\documentclass[sigconf, authorversion, balance]{acmart}

\usepackage[show]{chato-notes}
\usepackage{graphicx}
\usepackage{placeins}
\usepackage{enumitem}
\usepackage{kantlipsum}
\usepackage{multirow}
\usepackage{subfig}
\usepackage{makecell}
\usepackage{babel}
\usepackage{float}
\usepackage{wrapfig}
\usepackage[para]{footmisc}

\usepackage[normalem]{ulem}
\useunder{\uline}{\ul}{}

\def\bertrec{BERT\-4\-Rec}

\usepackage{array}
\newcolumntype{?}{!{\vrule width 1pt}}
\def\hlinewd#1{%
\noalign{\ifnum0=`}\fi\hrule \@height #1 %
\futurelet\reserved@a\@xhline} 

\usepackage{booktabs,caption}
\usepackage[flushleft]{threeparttable}

\usepackage{ragged2e}
\usepackage{multicol}
\graphicspath{ {./images/} }

\usepackage{natbib}

\MakeRobust{\ref}%

\makeatletter
\newcommand{\labeltext}[2]{%
  \@bsphack
  \csname phantomsection\endcsname %
  \def\@currentlabel{#1}{\label{#2}}%
  \@esphack
}
\makeatother

\usepackage{algorithm}%
\usepackage{algpseudocode}%
\algrenewcommand\algorithmicrequire{\textbf{Input:}}
\algrenewcommand\algorithmicensure{\textbf{Output:}}

\usepackage{pifont}%

\usepackage{svg}

\definecolor{smcolor}{rgb}{0.7, 0.4, 0}

\newcommand{\cms}[1]{\textcolor{black}{#1}}
\newcommand{\sm}[1]{\textcolor{black}{#1}}
\newcommand{\craig}[1]{\textcolor{black}{#1}}
\newcommand{\sasha}[1]{\textcolor{black}{#1}}
\newcommand{\srs}[1]{\textcolor{black}{#1}}
\newcommand{\crs}[1]{\textcolor{black}{#1}}

\newcommand{\scr}[1]{\textcolor{black}{#1}}

\def\maxPct{\tau}

\copyrightyear{2022}
\acmYear{2022}
\setcopyright{acmcopyright}\acmConference[RecSys '22]{Sixteenth ACM Conference on Recommender Systems}{September 18--23, 2022}{Seattle, WA, USA}
\acmBooktitle{Sixteenth ACM Conference on Recommender Systems (RecSys '22), September 18--23, 2022, Seattle, WA, USA}
\acmPrice{15.00}
\acmDOI{10.1145/3523227.3546785}
\acmISBN{978-1-4503-9278-5/22/09}

\begin{document}

\title{Effective and Efficient Training for Sequential Recommendation using Recency Sampling}

 \author{Aleksandr Petrov}
 \affiliation{%
   \institution{University of Glasgow} \country{United Kingdom}}

 \email{a.petrov.1@research.gla.ac.uk}

 \author{Craig Macdonald}
 \affiliation{%
   \institution{University of Glasgow} \country{United Kingdom}}
 \email{craig.macdonald@glasgow.ac.uk}

\begin{abstract}
Many modern sequential recommender systems use deep neural networks, which can effectively estimate the relevance of items but require a lot of time to train. Slow training increases expenses, hinders product development timescales and prevents the model from being regularly updated to adapt to changing user preferences. Training such sequential models involves appropriately sampling past user interactions to create a realistic training objective. The existing training objectives have limitations. For instance, next item prediction never uses the beginning of the sequence as a learning target, thereby potentially discarding valuable data. On the other hand, the item masking used by \bertrec{} is only weakly related to the goal of the sequential recommendation; therefore, it requires much more time to obtain an effective model. Hence, we propose a novel Recency-based Sampling of Sequences training objective that addresses both limitations. We apply our method to various recent and state-of-the-art model architectures --  such as GRU4Rec, Caser, and SASRec. \sm{We show that the models enhanced with our method can achieve performances exceeding or very close to state-of-the-art \bertrec{}, but with much less training time.}

\end{abstract}

\settopmatter{printfolios=true}
\maketitle

\section{Introduction}

\emph{Sequential recommender models} is a class of recommender systems,  which consider the order of the user-item interactions, are increasingly popular~\cite{quadrana2018sequence}.  Early sequential recommender systems used Markov Chains~\cite{rendle2010fpmc,zimdars2001temporal}, \craig{however most} modern \sm{ones} use deep neural networks, and have adapted ideas from other domains such as language modelling~\cite{hidasi2015gru4rec, hidasi2018gru4recplus, kang2018sasrec, sun2019bert4rec} or image processing~\cite{tang2018caser}. These deep neural models \craig{have been shown to outperform} traditional \sasha{non-neural} methods by a \sasha{significant margin~\cite{sun2019bert4rec,kang2018sasrec,yuan2019nextitnet,tang2018caser}.}

\begin{figure}
\centering
\includesvg[width=0.5\textwidth]{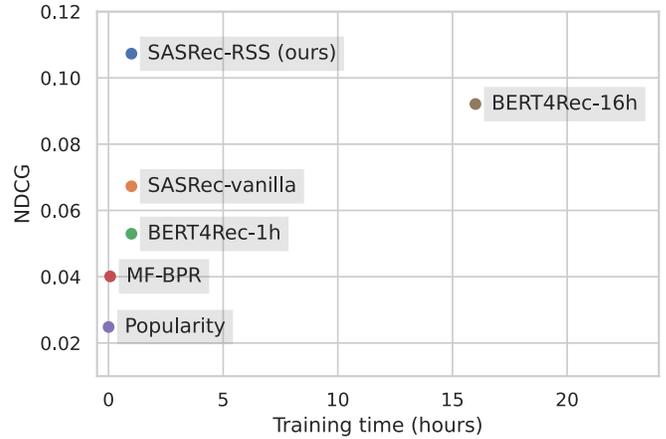}
  \caption{The SASRec~\cite{kang2018sasrec} model trained with our \craig{proposed} training method outperforms \bertrec{} on the MovieLens-20M dataset~\cite{harper2015movielens} and requires much less training time. SASRec-vanilla corresponds to the original version of SASRec; BERT4Rec-1h and BERT4Rec-1h are versions of BERT4Rec trained for 1 hour and 16 hours respectively.}\label{fig:motivation}
\end{figure}

 However, the most advanced sequential models, such as \bertrec{}, suffer from the slow training problem. Indeed, \sasha{our experiments show that in order to reproduce result reported in the original publication, \bertrec{} requires more than 10 hours training using modern hardware \scr{(see also replicability paper~\cite{petrov2022replicability})}}. This is illustrated in Figure~\ref{fig:motivation}, which portrays the NDCG@10 of MF-BPR~\cite{rendle2009bpr}, SASRec~\cite{kang2018sasrec} and \bertrec{}~\cite{sun2019bert4rec} models for different training durations \sasha{on the MovieLens-20M dataset~\cite{harper2015movielens}}. 

\hyphenation{Further-more}

Slow training is a problem in both research and production environments. For research, slow training limits the number of experiments that can run using available computational resources. In production, it increases the costs of using recommender systems due to the high running costs of GPU \sasha{or TPU} accelerators. Furthermore, slow training hinders how quickly the model can be retrained to adapt to changing user interests. For example, when a new episode of a popular TV show is released, the recommender system might \sasha{still be recommending} the old episode because it was not retrained yet. Hence, in this paper, we focus on the time-limited training of models. The main question we address in this paper is \emph{can the training of existing sequential recommendation models be improved so that they \craig{attain} state-of-the-art performance in limited training time?}

\craig{The primary components of model training can be characterized as follows: (i) the model architecture that is being \cms{trained}, (ii) the training objective that defines \cms{what} the model is being trained to reconstruct, and (iii) the loss function used to measure its success. Although all three components have a marked impact on training efficiency, in this work, we focus on the training objective, identifying two key limitations in existing approaches, \sm{as well as an appropriate loss function for the objective.}}

\craig{Among the training objectives in the literature, sequence cont\-inuat\-ion}~\cite{hidasi2015gru4rec, hidasi2018gru4recplus, tang2018caser} (including its popular form, next item prediction) is probably the most intuitive and popular. However, this objective never uses the beginning of the sequence as a training target\craig{, hence} \cms{it} discards potentially valuable knowledge and limits the number of training samples it can generate from a single sequence.

Second, in the \craig{item} masking approach -- which is used by BERT\-4\-Rec~\cite{sun2019bert4rec} -- the task of the model is to recover masked items at any position in the sequence, which is a much more general and complex task than \sm{the} next item prediction. \craig{We argue that this is only weakly related to the end goal of sequential recommendation.} Indeed, we will show that, despite leading to better results, the more general training task requires \craig{considerable} training time.

\craig{These limitations of the} existing approaches \cms{motivate} us to design a new  \emph{Recency-based Sampling  of Sequences (RSS)} approach that probabilistically selects positives from the sequence to build training samples. In our method, more recent interactions have more chances of being sampled as positives; however, due to the sampling process' probabilistic nature, even the oldest interactions have a non-zero probability of being selected as positives.

\craig{Our experiments are conducted on four large sequential recom\-mender datasets, and demonstrate the application of the proposed RSS approach upon three recent sequential recommendation model architectures (GRU4Rec, Caser and SASRec), \cms{when} combined with both pointwise and listwise loss functions.} We find that RSS improves the effectiveness of all three model architectures. Moreover, on \srs{all four experimental} datasets, \srs{versions} of RSS-enhanced SASRec trained for one hour can markedly outperform state-of-the-art baselines. Indeed, RSS applied to the SASRec model can result in an \sasha{60\%} improvement in NDCG over a vanilla SASRec, and a \sasha{16\%} improvement over a fully-trained BERT4Rec model, despite taking \sasha{93\%} less training time than BERT4Rec (see also Figure~\ref{fig:motivation}).

In short, the main contributions of this paper are as follows: (i) We identify limitations in the existing training objectives used by sequential recommendation models; (ii) We propose Recency-based Sampling  of Sequences, which emphasises the importance of more recent interactions during training; (iii) \cms{We perform extensive empirical evaluations} on four sequential recommendation datasets, demonstrating significant improvements over existing state-of-the-art approaches. The structure of this paper is  as follows: Section~\ref{sec:related} provides \crs{a background in} sequential reco\-m\-m\-endation; Section~\ref{section:training} covers existing approaches and identifies their limitations; \craig{In Section}~\ref{sec:rss} we explain Recency-based Sampling of Sequences for efficient training. Section~\ref{sec:expsetup} describes research questions and experimental setup; \craig{In Sections}~\ref{section:results} \&~\ref{sec:conc} we \craig{respectively} provide analysis of the experiments and \craig{concluding remarks}.

\section{Background}\label{sec:related}
In the following, we provide an overview of neural sequential recommendation models. Indeed, over the last several years, most of the next item prediction \cms{approaches} have \craig{applied} deep neural network models. Some of the first solutions based on deep neural networks were GRU4rec~\cite{hidasi2015gru4rec} and the improved  GRU4Rec\textsuperscript{+}~\cite{hidasi2018gru4recplus} (\sasha{using an improved listwise loss function}), \craig{which are} models that use the Recurrent Neural Networks (RNN) architecture. \craig{On the other hand}, Caser~\cite{tang2018caser} uses ideas from computer vision; it generates a 2D ``image" of the sequence using item embeddings and then applies horizontal and vertical convolution operations to that image. Another model that is based on \craig{convolution} operation is NextItNet~\cite{yuan2019nextitnet}, \craig{which} applies several layers of 1D convolutions to generate rich semantic representations of \craig{each user} sequence. \sasha{These models all use variations of \craig{a} sequence continuation task for training, details of which we provide in the Section~\ref{section:training}}.

\begin{figure}
    \centering
    \includesvg[width=0.4\textwidth]{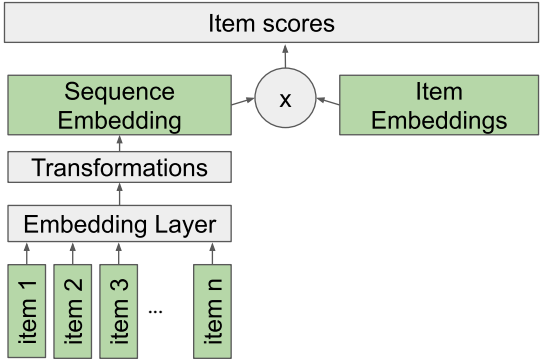}
    \caption{Principal architecture of many sequential recommenders. \craig{This applies to} GRU4rec~\cite{hidasi2015gru4rec}, GRU4rec\textsuperscript{+}~\cite{hidasi2018gru4recplus}, Caser~\cite{tang2018caser} and, with minor modifications to SASRec~\cite{kang2018sasrec}.}
    \label{fig:principle_architecture}
\end{figure}

\craig{Figure~\ref{fig:principle_architecture} illustrates the principal architecture of many of the sequential recommendation models used in this work.} These generate an \cms{embedding} of the user's sequence and then multiply this embedding by the matrix of item embeddings to \craig{obtain} item scores. GRU4rec,  Caser, and --  with minor modifications (see Section~\ref{section:training}) -- SASRec use this architecture. Recent state-of-the-art sequential recommendation models use variations of the transformer~\cite{vaswani2017attention} architecture. SASRec~\cite{kang2018sasrec} uses transformer blocks to predict the next item in the sequence based on all previous elements. \bertrec{}~\cite{sun2019bert4rec} adapts the well-\cms{known} BERT language model~\cite{devlin2018bert} for the sequential recommendation task.  Following the original \craig{BERT} model, \bertrec{} \craig{is trained} to reconstruct {\em masked} items that are hidden from the model during training. In particular, as both SASRec and \bertrec{} use \crs{the} transformer architecture, the only significant difference between these two models is the training scheme. Using item masking, \bertrec{} outperforms SASRec in terms of quality; however, it requires much more training time. \craig{In this work, we identify limitations in the existing training objectives,} which we discuss further in Section~\ref{section:training}.  Indeed, the goal of this work is to close the gap between effectiveness and efficiency and design a new training scheme that allows \sm{matching} the performance of state-of-the-art models within limited training time.

Finally, recent advances have used graph neural networks (GNNs) for sequential recom\-mendation~\cite{Graph2020gag, qiu2020GraphExploiting, qiu2021GraphExploiting, qiu2021GraphMemory}. \sasha{These models usually use additional information, such as cross-session connections or item attributes. \cms{In this work, we focus on a more general case of sequential recommendations, without the assumption of availability of cross-session (user) information or cross-item connections; therefore graph-based models, as well as those tailored for personalised shopping basket completion (e.g.~\cite{meng2021vbcar, wan2018representing}) are out of the scope of this work.}} %
On the other hand, CL4SRec~\cite{xie2020contrastive} applies data augmentation by modifying the input sequences (e.g.\ cropping, masking, or reorder\-ing). These augmentations are orthogonal to changing the training task and could be used together with an improved training objective. \craig{Nevertheless, we focus on the training objective for sequential models operating without use of GNNs nor data augmentation. We provide details of these training objectives in the next section.}

\section{Training Sequential Recommendation Models}\label{section:training}

Consider a set of users $U$ and items $I$. Each user $u \in U$ has a sequence of interactions $s_u = \{i_{u_1}, i_{u_2}, i_{u_3} ... i_{u_n}\}$ where items $i_{u_\tau} \in I$ are ordered by the interaction  time. The \emph{next item prediction task} is defined as follows: given \craig{a sequence} $s_{u}$, rank the items from $I$, according to their likelihood of being the sequence continuation~$i_{u_{n+1}}$. This task corresponds to \textit{Leave One Out} evaluation - hold out the last element from each user's interactions sequence and then evaluate how well a model can predict each held-out element.

\sasha{As mentioned in Section~\ref{sec:related}, the best models for the next item prediction task are based on deep neural networks.} Generally speaking, their training procedure consists of iterations of the following steps: 
(1) Generate a batch of training samples, each with positive and negative items; (2) Generate predictions, using our model; (3) Compute the value of the loss function; (4) Update model parameters using backpropagation.

We aim to improve the training of existing models, so step~2 is \craig{not within} the scope of our work. \craig{Backpropagation} (step~4) -- e.g. through stochastic gradient descent -- is a very general and well-studied procedure, and we follow the best practices used by the deep learning models, details of which we describe in \craig{Section}~\ref{sec:expsetup}. This leaves us with two essential parts of model optimization - generation of the training samples and the loss function. \sasha{These two parts are not independent: a loss function designed for one training task does not always fit into another. For example, BPR-max loss (used by GRU4Rec\textsuperscript{+}~\cite{hidasi2018gru4recplus}) has an assumption of only one positive item per training sample and therefore is not applicable to a sequence continuation with multiple positives task, as used by Caser. Hence, a new training task requires selection of an appropriate loss function. We further discuss some possible choices of the loss functions for our proposed method \craig{later} in Section~\ref{section:loss}.} In the \craig{following}, we review approaches to generate training samples and identify their limitations, a summary of which we provide in the Section~\ref{section:summary_of_limitations}. %

\hyphenation{con-ti-nuation}
\hyphenation{re-construct-tion}
\subsection{Generation of Training Samples}\label{section:generating}
\begin{figure*}
      \includesvg[width=\linewidth]{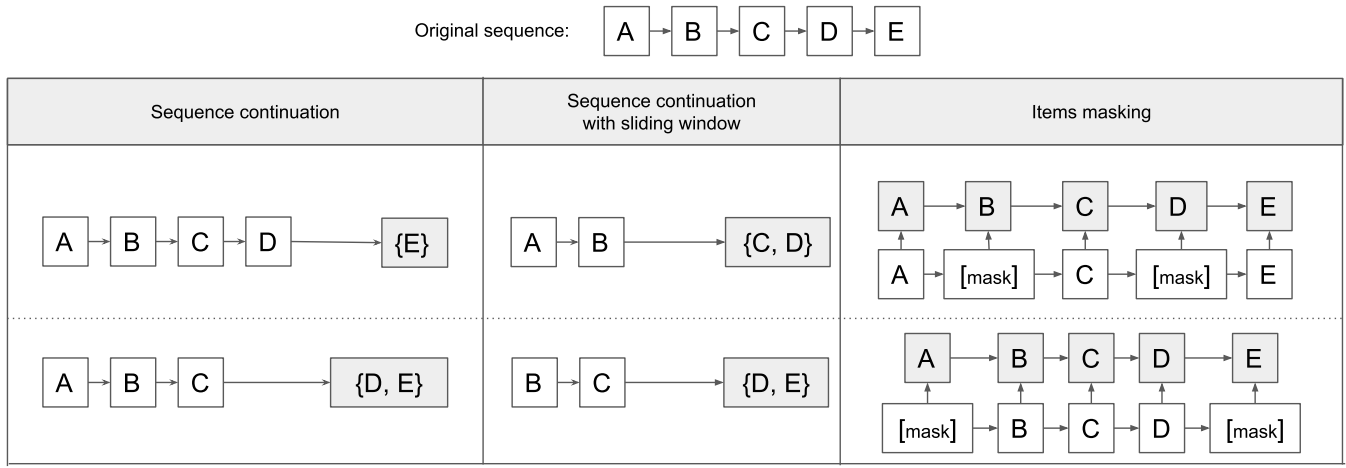}
  \caption{Training samples generation strategies used in existing models. White boxes represent model inputs and filled boxes represent model outputs.
In Sequence Continuation, the sequence is split into two parts, with the aim of predicting whether or not an item belongs to the second part based on the sequence of elements in the first part. In Sequence Continuation with a sliding window, we first generate shorter sub-sequences from the original sequence and then apply the sequence continuation method. In item masking, some elements are removed and replaced  with a special "[mask]" value, with the aim of correctly reconstructing these masked items.}\label{fig:sampling_stragies}
\end{figure*}

A training sample for a sequential model consists of three parts - the input sequence, positive items, and negative samples. Sequential recommender models~\cite{kang2018sasrec, hidasi2015gru4rec, hidasi2018gru4recplus, tang2018caser} treat ground truth relevance as a binary function; by definition, every non-positive item is negative. In practice, to \craig{make the training more tractable, most} models only consider samples of negative items, identified using \craig{techniques} such as random sampling~\cite{kang2018sasrec, rendle2009bpr}, in-batch negatives~\cite{hidasi2015gru4rec}, or the negatives with highest scores~\cite{yuan2016lambdafm}. This work focuses on constructing positive samples. Negatives sampling \craig{approaches} are orthogonal to positive sampling and can be applied independently. We do not use negative sampling in our work and leave improvement of our method via negative sampling to  future research.
In \craig{the remainder of this section}, we describe positive sampling strategies for sequential recommendations. Figure~\ref{fig:sampling_stragies} illustrates sequence continuation and item masking, the most commonly used strategies, \craig{which we discuss in turn below.}

Matrix factorization methods use a straight\-forward \emph{matrix reconstruction} training objective: for each user $u$ and item $i$, the goal of the model is to estimate whether the \sm{user} interacted with the item. This goal leads to a simple training samples generation procedure - we sample (user, item) pairs as inputs and assign labels for the pairs based on interactions. A classic model that uses matrix reconstruction is Bayesian Personalized Rank (BPR)~\cite{rendle2009bpr}, which we use as one of our baselines.  The main disadvantage of matrix reconstruction is that it does not consider the order of the interactions, and therefore sequential recommendation models can not use it.

\cms{In the \emph{sequence continuation} training objective,  training samples are generated by} splitting the sequence of interactions into two consequent parts:
\[
       s = \{i_1, i_2, i_3 .. i_n\} \mapsto \begin{cases} s_{input} = \{i_1, i_2, i_3, ... i_{n-k}\} ; \\
            s_{target} = \{i_{n-k+1}, i_{n-k+2}, .. i_{n}\}
            \end{cases}
\]
where $k$ is a hyperparameter. 
We use $s_{input}$ as the model input, and assign label $1$ to the postive items from $i_{+} \in s_{target}$ and label $0$ to the negative items $i_{-} \notin s_{target}$. If $k$ is equal to 1, the sequence continuation task turns into the next item prediction task, which matches the end goal of sequential recommender systems. 

\hyphenation{pre-fe-re-nces}
Using sequence continuation in its basic form, we can produce precisely one training sample out of a single sequence of interactions. Some models (e.g.\  Caser~\cite{tang2018caser}) use the sliding window approach to generate more than one sequence - which generates shorter subsequences out of a whole sequence and then creates training samples out of these shorter subsequences. The sliding window approach allows to generate up to $n - 1$ training samples out of a sequence of $n$ interactions. However, working with shorter sequences limits the model by modeling only short-term user preferences, and the models have to find a balance between the number of generated samples and the maximum length of the sequence~\cite{tang2018caser}. GRU4rec, GRU4rec\textsuperscript{+} and Caser models use variations of the sequence continuation task for training. %
The training task used by SASRec \sasha{and NextItNet~\cite{yuan2019nextitnet}} is slightly different: \sasha{they} work as a sequence-to-sequence models where the target sequence is shifted by one element compared to the input. These models predict the second element of the input sequence by the first, third by the first two, etc.. By design of the models, when they predict the $j^{th}$ item in the output, \sm{they only have} access to the first $(j-1)$ elements of the input so that this shifted sequence prediction task essentially is $n$ independent sequence continuation tasks.

Thus, the main limitation of sequence continuation is that it only generates a small number of training samples out of a single sequence and the  items in the first part of the user's sequence never have a chance to be selected as a target, which means that the recommender system is unlikely to learn how to recommend these items, even though they may be relevant for some users. We refer to this limitation as Limitation \ref{limitation:sampleperseq}.

\hyphenation{pre-ferences}

\bertrec{}~\cite{sun2019bert4rec} \srs{uses an \emph{item masking} training objective}, which it inherited from the original BERT model. In BERT, the idea is to mask some \craig{terms} from the sentence and then ask the model to reconstruct these hidden elements. Similarly, in \bertrec{}, some items in the sequence are masked, and the model is retrained to recover these items. The target sequence, in this case, exactly matches the original sequence \sm{(without masking):}
$$
       s = \{i_1, i_2, i_3, i_4, .. i_n\} \mapsto \begin{cases}  s_{input} = \{i_1, [mask], i_3, [mask], ... i_{n}\}; \\ s_{target} = \{i_1, i_2, i_3, i_4, .. i_n\} 
       \end{cases}
$$

\hyphenation{compa-red}
 This \cms{approach generates} up to $2^n$ training samples out of a single training sequence of length $n$. \bertrec{} does not mask more than $\maxPct$ percent \craig{of} items in \craig{a} sequence, where $\maxPct$ is a hyperparameter; however, it still generates many more training samples compared to the single training sample generated \craig{from a sequence under sequence continuation}. As Sun et al.~\cite{sun2019bert4rec} \craig{showed}, more training \craig{ensures to avoid} overfitting and \craig{achieves} better performance compared to other models with similar architecture. 

\craig{However, we argue that the} main disadvantage of the item masking approach is that it is weakly related to the next item prediction task. To make \craig{a} prediction, \bertrec{} adds the $[mask]$ element to the end of the input sequence and tries to reconstruct it; so that training and evaluation samples have a different distribution. The model must learn how to solve the evaluation task (reconstruct the last item in the sequence) as part of a much more general and more complicated task (reconstruct any item in the sequence). \bertrec{} adds a small proportion of training samples with only the last element masked to address this mismatch, but \craig{the consequence is} still a substantially more complicated training task and longer time to converge compared to the models that use sequence cont\-in\-ua\-t\-ion. \sasha{We refer to this problem of weak \craig{correspondence} to the original task as Limitation \ref{limitation:weakrelation}}.

\subsection{Summary of Limitations}
\label{section:summary_of_limitations}
We reviewed two main training \sm{objectives} used in sequential reco\-m\-m\-end\-at\-ions - sequence continuation (including its variations, shifted sequence prediction, and sliding window) and item masking. Indeed, as argued above, both of these training objectives have their limitations, which we summarize as follows
\begin{enumerate}[font={\bfseries}, label={L\arabic*}]
\item \label{limitation:sampleperseq} Sequence continuation \cms{can only generate} a small number of training samples \craig{from a} single training sequence. This allows training to be performed relatively quickly, but performance of these models is lower compared to \craig{a} state-of-the-art model such as \bertrec{}. 

\item \label{limitation:weakrelation}  Reconstruction of masked items is a very general task, which is loosely connected to the sequential recommendation task. Using this task, models can reach state-of-the-art performance, \cms{but model training can take markedly longer than other training objectives.}
\end{enumerate}

 In the next section, we introduce \emph{Recency-based Sampling of Sequences}, a novel training task that \craig{addresses} these \craig{limitations} \sasha{and discuss possible choices of the loss function for this training task}. %

\section{Recency\mbox{-}based~Sampling~of~Sequences} \label{sec:rss}
\sasha{As shown in Section~\ref{section:training}, to train a model we need to have a training task and choose a loss function that matches the task. Hence, the training task and the loss function are both essential parts of our solution.} \craig{In this section we introduce both training objective (Section~\ref{ssec:rss:object}) and choice of loss function (Section~\ref{section:loss})}.

Recency-based Sampling of Sequences (RSS) is a training objective that is closely related to the sequential recommendations and allows to generate many training samples out of a single user sequence simultaneously. To \craig{address the limitations of existing training objectives} described in Section~\ref{section:summary_of_limitations}, we first outline the principles used to design our training task:
\begin{enumerate}[font={\bfseries}, label={P\arabic*}]
    \item \label{principle:each_item}
    Each element in a sequence can be selected as the target; multiple items can be selected as a target in each training sample. Using this principle, we match the main advantage of the item masking approach - generating up to $2^n$ training samples out of each user sequence. \sasha{This principle addresses  Limitation~\ref{limitation:sampleperseq}}.
    
    \hyphenation{advanta-ges}
    \item \label{principle:recency} More recent training interactions in a sequence better indicate the user's interests, and hence these are more realistic targets. User interests change over time, and one of the main advantages of sequential recommender systems is taking these changes into account. Therefore, the methods that rely on this principle will retain a close connection to sequential recommendations. \sasha{This principle addresses Limitation~\ref{limitation:weakrelation}}.
\end{enumerate}

 In our proposed training objective, to follow these two principles, we use a \emph{recency importance function}, $f(k)$, that is defined for each position $0$ .. $n-1$ in the sequence of the length $n$ and indicates chances of each position to be selected as a target:  probability of an item at position $k$ of being selected as a positive is proportional to the value of $f(k)$.  \cms{$f(k)$} \crs{must} \craig{exhibit the following properties}: $f(k)$ is positive \crs{($f(k) > 0$) and \craig{monotonically} growing ($f(k) \leq f(k+1)$).} 
This first property corresponds to \sasha{Principle \ref{principle:each_item}} and \sm{defines that the likelihood of each item} to be selected as \sm{a target are} positive. \sm{The} second property corresponds to \sasha{Principle \ref{principle:recency}}, and \sm{ensures that} more recent items have higher or equal chances to be selected as \craig{a} target.

\def\cnt{c}

To generate a training sample, we first calculate $c$ - how many target items we want to sample. Following \bertrec{}, we define a parameter \sasha{$\maxPct$  that controls the maximum percentage of items that can be used as targets and then calculate $c$ via multiplying $\maxPct$ by the length of the sequence}. We then randomly sample, with replacement, $c$ targets from the sequence, with the probability of being sampled, $p(i)$ proportional to the value of a recency importance function, $f(i)$:
\begin{align}
    p(i) = \frac{f(i)}{\sum_{j=0}^{n-1}f(j)}
\end{align}
    
\begin{figure}
    \centering
    \includesvg[width=0.45\textwidth]{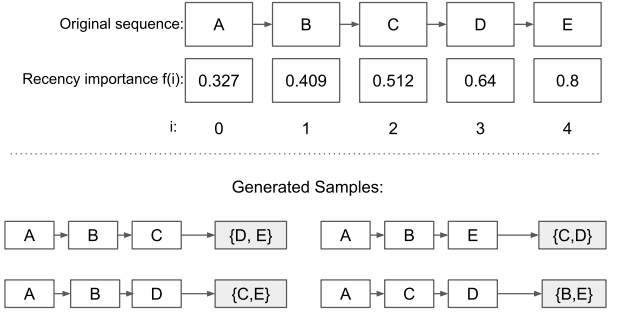}
    
  \caption{Recency-based Sampling of Sequences. The beginnings of the sequences remains largely unchanged, whereas elements from the end of the sequence are chosen as positive samples \cms{more} frequently.}\label{fig:generated_samples}
 \end{figure}

We generate the input sequence to the model by removing targets from the original sequence. The full procedure is described in Algorithm~\ref{alg:ALG1}. One example of a \sasha{recency importance function that has the required properties} is the exponential function: 
\begin{align}\label{eq:exponential_importance}
f(k) = \alpha^{n-k}
\end{align}
\craig{where} $0 < \alpha \leq 1$ is a parameter that controls importance of the recent items in the sequence and $n$ is the sequence length. If $\alpha=1$, then each item has equal chances of being sampled as a target, and Recency-based Sampling of Sequences to the item masking approach \sm{(but without providing the positions of masked items) or to the matrix reconstruction approach,} where items are sampled uniformly from the sequence. If $\alpha$ is close to zero, items from the end of the sequence have a much higher chance of being sampled, and therefore \sm{RSS} becomes equivalent \craig{to} the sequence continuation task. Figure~\ref{fig:generated_samples} provides an example of the recency importances (for $\alpha=0.8$) and the generated samples.

\subsection{The RSS Training Objective}\label{ssec:rss:object}

 There are other possible position importance functions, such as linear $f(k) = k + 1$, and the best function may be a property of particular dataset. \cms{However}, in this work, we only consider the exponential function, leaving other possibilities \craig{to future} research.

\subsection{Loss Functions for RSS}\label{section:loss}
The second important component of the training procedure is the loss function. Loss functions for recommender systems can be generally divided into three categories - \emph{pointwise} (optimize the relevance estimation of each item independently), \emph{pairwise} (optimize a partial ordering between pairs of items) and \emph{listwise} (optimize the recommendations list as a whole)  losses~\cite{Liu09ftir}. RSS works with all types of loss functions that support multiple positive samples \craig{within each} training sample.

\hyphenation{hyper-parameter}

GRU4rec\textsuperscript{+}~\cite{hidasi2018gru4recplus} showed the advantages of applying a listwise loss function above pointwise and pairwise methods, however the Top-1-max and BPR-max losses introduced in that paper have an assumption that there is only one positive item \craig{within each} training sample. 
Instead, we use LambdaRank~\cite{burges2010ranknet} (or $\lambda$Rank), another listwise optimization loss function. \craig{$\lambda$Rank has been widely deployed in training learning-to-rank scenarios~\cite{chapelle2011yahoo, qin2020neural} for web search.  Similarly, $\lambda$Rank has been shown to be advantageous for recommender tasks~\cite{li2021new}, for example when applied to Factorisation Machines~\cite{yuan2016lambdafm} or transformer-based sequential models~\cite{petrov2021booking}.}
$\lambda$Rank~\cite{burges2010ranknet} uses \textit{Lambda Gradients} instead of objective function gradients in the gradient descent method. The Lambda Gradient for an item $i \in I$ is defined as follows: 
\begin{align}
    \lambda_i = \sum_{j \in I}|\Delta NDCG_{ij}|\frac{-\sigma}{1 +  e^{\sigma(s_i - s_j)}}\label{eqn:lrank}
\end{align}
where $s_{i}$ and $s_{j}$ are predicted scores, $\Delta NDCG_{ij}$ is a change in NDCG metric in case of swapping items $i$ \& $j$, \sm{and $\sigma$ is a hyperparameter typically set to 1}. 

In addition to $\lambda$Rank, we also \sasha{experiment with} the standard \sasha{Binary Cross-Entropy (BCE), following~\cite{tang2018caser, kang2018sasrec}, to investigate the effect of the listwise loss and the necessity of \crs{both training objective and the loss function in our solution.}}

\section{Experiment Setup}\label{sec:expsetup}
\craig{In the following, we list our research questions (Section~\ref{ssec:expsetup:rq}), \crs{our experimental datasets} (Section~\ref{ssec:expsetup:datasets}), the recommender models on which we build and our comparative baselines (Section~\ref{ssec:expsetup:models}), and finally evaluation details (Section~\ref{ssec:expsetup:eval})}.

\hyphenation{exponen-tial}
\subsection{Research Questions}\label{ssec:expsetup:rq}
\craig{Our} experiments \craig{aim to address} \craig{the} following research questions:
\begin{enumerate}[font={\bfseries}, label={RQ\arabic*}, wide, labelwidth=!, labelindent=0pt]
    \item Does Recency-based Sampling of Sequences (RSS) help for training sequential recommendation models compared to sequence continuation? \label{rq:improves}
    \item Does a listwise $\lambda$Rank loss function benefit RSS training? \label{rq:loss}
    \item What is the impact of the recency \sm{importance} parameter \sm{$\alpha$  in the exponential recency importance function (Equation~\eqref{eq:exponential_importance})} of RSS? \label{rq:recency}
    \item How do RSS-enhanced models compare with state-of-the-art baselines? \label{rq:sota}
\end{enumerate}

 \begin{algorithm}[tb]
        \caption{Recency-based Sampling of Sequences}
        \label{alg:RSS}
        \begin{algorithmic}
        \Require $sequence$ - a sequence of interactions; $\maxPct$ - maximum percent of target items; $f$ - recency importance function
        \Ensure $input$ is a generated input sequence for the model; $target$ is a set of sampled positive items
        \Function {RecencySequenceSampling}{$sequence$, $\maxPct$, $f$}
            \State $sampledIdx \gets set()$
            \State $n \gets length(sequence)$; $\cnt \gets max(1, int(n*\maxPct)))$        
            \State $prob \gets Array[n]$
            \State $prob[i] \gets  \frac{f(i)}{\sum_{j=0}^{n-1} f(j)}$ \textbf{for} i \textbf{in} [$0$, $n-1$]
            
            \State $sampledIdx \gets random.choice(range(0 .. n-1), \cnt, prob)$
            \State $input \gets  list()$; $target \gets set()$
            \For{$i \gets 0, n-1$}
                \State  \textbf{if} $i \in sampledIdx$ \textbf{then} $target.add(sequence[i])$ \textbf{else} $input.append(sequence[i])$
            \EndFor
            \State \textbf{return} $input, target$
        \EndFunction
        \end{algorithmic}\label{alg:ALG1}
        
        \noindent\rule{\linewidth}{0.5pt}
         \justifying \footnotesize We assume that function $random.choice(a, \cnt, p)$ is an equivalent of the $numpy.random.choice$ function from the numpy package. It iteratively samples $\cnt$  samples from collection $a$, where the probability of each item $i$ of being sampled equals $p[i]$ at each stage, with replacement.
\end{algorithm}

\subsection{Datasets}\label{ssec:expsetup:datasets}
\craig{Our experiments are performed on \crs{four} large-scale datasets for sequential recommendation:}

    \textit{MovieLens-20M}~\cite{harper2015movielens} is a \cms{movie} recommendation dataset, and is popular for benchmarking sequential recom\-menders~\cite{sun2019bert4rec, fischer2020kebert, ma2019hierarchical, qiu2021GraphMemory, ma2020disentangled, zhao2021adversarial, li2021intention, cho2020meantime, wu2021seq2bubbles}. \srs{Note that MovieLens-20M is a ratings dataset, where users rate movies with stars, however following common practice~\cite{sun2019bert4rec, ma2020disentangled, li2021intention} we consider all ratings as positive interactions. \crs{However, MovieLens-20M timestamps correspond to the time when ratings were provided rather than when the items were} consumed, so the task is best described as ``next movie to rate'' rather than ``next movie to watch''. \crs{Nevertheless}, as versions of the MovieLens dataset are used in both well-cited~\cite{kang2018sasrec, sun2019bert4rec, huang2018improving} and recent~\cite{qiu2022contrastive, zhang2022dynamic, zhan2022transrecplusplus} sequential recommendation papers, we conclude that it is well suited for the problem and it is important to include it as one of our benchmarks.} 
    
    \textit{Yelp}\footnote{\href{https://www.yelp.com/dataset}{https://www.yelp.com/dataset}} - is a businesses reviews dataset. It is another popular dataset for sequential recommendations~\cite{zhou2020s3, amjadi2021katrec, bian2021contrastive, wang2022sequential, qiu2022contrastive, padungkiatwattana2022arerec}. As for MovieLens-20M, we consider all user reviews as positives.
    
    \textit{Gowalla}~\cite{cho2011gowalla} contains \craig{user checkins to a location-based social network}. \craig{This dataset contains a} large number of items (more than~$10^6$) and is very sparse: \sasha{it has only 0.0058\% out of all possible user-item interactions.}
    
    \textit{Booking.com~}\cite{goldenberg2021booking} \craig{is a travel} \cms{destination} dataset. Each \craig{interaction} sequence in this dataset represents a single multi-city trip of a user. In contrast to other types of recommendations, such as movies or books, multi-city trips have \craig{a} strong sequential nature. Indeed, for example, if a user is \sm{making} a road trip by car, there could be only one or two \sm{neighboring} cities where the user can stop, and \craig{hence all other more distant items are non-relevant.} %
    This strong sequential nature could be problematic \sasha{for RSS, as it contradicts Principle \ref{principle:each_item}, which says that any item in the sequence can be selected as a relevant target for the \sm{preceding} items}. %

\noindent \srs{Following common practice~\cite{sun2019bert4rec, zhou2020s3, kang2018sasrec}, we discard cold-start users with fewer than 5 interactions from each dataset.} Table~\ref{tab:datasets} reports the salient statistics of the three datasets.

\begin{table*}[tb] 
\caption{Datasets we use for experiments.}\label{tab:datasets}
\begin{tabular}{lrrrrrr}
\toprule
Name &  Users &  Items &  Interactions &  \makecell{Average\\length} &  \makecell{Median\\length} &  sparsity \\
\midrule
Booking.com &     140746 &      34742 &            917729 &             6.52 &                   6 &  0.999812 \\
Gowalla &      86168 &    1271638 &           6397903 &            74.24 &                  28 &  0.999942 \\
Yelp    &     287116 &     148523 &           4392169 &            15.29 &                   8 &  0.999897 \\
MovieLens-20M        &     138493 &      26744 &          20000263 &           144.413530 &                  68 &  0.994600 \\
\bottomrule
\end{tabular}
\end{table*}

\subsection{Models}\label{ssec:expsetup:models}
\noindent \subsubsection{Experimental Architectures}
We \craig{experiment using RSS \cms{with}} three \craig{recent} model architectures for sequence recommendation: (i)~\textit{GRU4Rec}~\cite{hidasi2015gru4rec} is a sequential recommender architecture based on recurrent networks; (ii)~\textit{Caser}~\cite{tang2018caser} \craig{applies a convolutional neural network structure for sequential recommendation}. For our experiments, we use the basic architecture described in~\cite{yuan2019nextitnet}; (iii)~\textit{SASRec}~\cite{kang2018sasrec} is a sequential recommendation architecture based on transformers. \craig{The original} implementation of SASRec is trained as a sequence-to-sequence model, however only the \craig{final} element from the target sequence is used at inference time. In order to match our common training framework and train the model with the RSS training objective, we ignore all outputs of the architecture except the \craig{final} one. \srs{This is a notable change in the training process, because \crs{the} original SASRec computes its loss over all outputs. To make sure that this change does not lead to significant quality degradation we include the original version of SASRec as a baseline (see Section \ref{ssec:baselines}).}

We \cms{implement} %
these architectures using TensorFlow version 2~\cite{abadi2016tensorflow}. Note that for our experiments we \craig{reuse only the} architectures of these models and not the training methods or hyper-parameters. \craig{Indeed, because} our goal is to research the impact of the training task, the \craig{appropriate training parameters may differ from the original implementation.}

\sm{We implement RSS and sequence continuation training objectives on each of the three experimental architectures. We do not apply an item masking training objective with these architectures\srs{: item masking assumes that a model produces a scores distribution per masked item, which }is not compatible with those architectures; however, as discussed below, we include \bertrec{} as an item masking baseline.}

For our experiments, we set common training parameters for all model architectures, following the settings in~\cite{kang2018sasrec}. In particular, we \craig{limit the length} of the user sequences \craig{to} 50 items for all three datasets, and we set the size of the item embeddings to be 64. We use \craig{the Adam optimizer, applying} the default learning rate of 0.001 and $\beta_2$ parameter set to 0.98. \craig{Following \bertrec{}~\cite{sun2019bert4rec}, we set the maximum percentage of a sequence to sample, $\maxPct=0.2$.} Except where otherwise noted, we set the recency parameter $\alpha$ to 0.8. \crs{Finally, in} order to estimate \sm{the} performance of the models under limited training time, we \crs{fix} the training time of all models to 1 hour. Experiments are conducted using 16-cores of an AMD Ryzen 3975WX CPU, 128GB of memory and an Nvidia A6000 GPU.

\hyphenation{compa-rable}
\noindent  \subsubsection{Baselines}
\label{ssec:baselines}
In order to validate that using Recency-based Sampling of Sequences it is possible to achieve performance comparable with state-of-the-art \craig{recommender models}, we compare with a selection of popular and state-of-the-art recommenders. We use the following models \srs{non-neural models} as baselines: (i)~\textit{Popularity} - the most popular items in the datset; (ii) \textit{MF-BPR} - Matrix factorization with BPR Loss~\cite{rendle2009bpr}. We use the implementation of this recommendation model from the popular LightFM library~\cite{kula2015lightfm}.
    
We also use two Transformer-based models as state-of-the-art baselines: (i) \textit{SASRec-vanilla} - the original version of SASRec recommender~\cite{kang2018sasrec}, a transformer-based model that uses a shifted sequence task, described in Section~\ref{section:generating}. To make the comparison fair with the RSS-enhanced variant, we \cms{limit the} training time of this model \craig{to} 1 hour; (ii) \textit{\bertrec{}} is another transformer-based model~\cite{sun2019bert4rec} based on the BERT~\cite{devlin2018bert} architecture. \craig{\bertrec{} has been shown to outperform other traditional and neural architectures and has been used as a strong baseline in a number of recent works (e.g.~\cite{qin2020neural, liu2021augmenting, koopmann2021cobert, lee2021moi}).} 

We use two versions of this model: {\em \bertrec{}-1h} denotes where the training time of \bertrec{} is limited to 1 hour, to allow a fair comparison in a limited-time setting; {\em \bertrec{}-16h}, where training time is limited to 16 hours in order to compare \sm{the} performance of our approach with the state-of-the-art model. \sasha{The original \bertrec{} publication~\cite{sun2019bert4rec} does not report \sm{the} required amount of training, but we \cms{find} empirically that reproducing the reported results takes around 16 hours on our hardware}. %

In contrast with other baselines, BERT4Rec calculates a score distribution across all items in the catalog for each element in the sequence, whereas other baselines calculate a single distribution of scores per sequence. This means that BERT4Rec requires $O(N)$ more memory per training sample for storing output scores and ground truth labels compared to other baseline models. This makes training original implementation of BERT4Rec infeasible when a dataset has too many items. Indeed, the original BERT4Rec publication~\cite{sun2019bert4rec} only reports results on relatively small datasets with no more than 55000 items and our own attempts to train BERT4Rec on large Gowalla dataset with more than 1 Million items failed because of memory and storage issues (see also Section~\ref{ssec:sota}). Hence, we do not report BERT4Rec results for Gowalla and leave scaling BERT4Rec to datasets with large number of items for the future research.

\begin{table*}[tb]
\caption{Comparing sequence continuation with Recency-based Sampling of Sequences training objectives under limited training for various model architectures.}\label{table:contvsrss}
\resizebox{0.98\textwidth}{!}{

\begin{threeparttable}
\subfloat[Recall@10]{
\begin{tabular}{|cl|ll|ll|ll|ll|}
\hline
\multicolumn{2}{|l|}{} & \multicolumn{2}{l|}{MovieLens-20M} & \multicolumn{2}{l|}{Yelp} & \multicolumn{2}{l|}{Gowalla} & \multicolumn{2}{l|}{Booking.com} \\ \hline
\multicolumn{1}{|l|}{Architecture} & Loss & Cont & RSS & Cont & RSS & Cont & RSS & Cont & RSS \\ \hline
\multicolumn{1}{|c|}{\multirow{2}{*}{GRU4Rec}} & BCE & 0.0221† & \textbf{0.0354*} & 0.0075† & \textbf{0.0100*†} & \textbf{0.0026*} & 0.0005 & 0.4621 & \textbf{0.4962*} \\
\multicolumn{1}{|c|}{} & $\lambda$Rank & 0.0082 & \textbf{0.1544*†} & 0.0009 & \textbf{0.0045*} & 0.0068† & \textbf{0.0119*†} & \textbf{0.4780†} & \textbf{0.5084*†} \\ \hline
\multicolumn{1}{|c|}{\multirow{2}{*}{Caser}} & BCE & 0.1424† & \textbf{0.1866*} & 0.0046† & \textbf{0.0099*†} & 0.0076 & \textbf{0.0081} & \textbf{0.5600*†} & 0.5454† \\
\multicolumn{1}{|c|}{} & $\lambda$Rank & 0.0330 & \textbf{0.1496*†} & 0.0009 & \textbf{0.0017*} & 0.0087† & \textbf{0.0157*†} & 0.4968 & \textbf{0.5273*} \\ \hline
\multicolumn{1}{|c|}{\multirow{2}{*}{SASRec}} & BCE & 0.1537† & \textbf{0.1888*} & 0.0146† & \textbf{0.0269*†} & 0.0089 & 0.0089 & \textbf{0.5845*†} & 0.5178 \\
\multicolumn{1}{|c|}{} & $\lambda$Rank & 0.1050 & \textbf{0.1968*†} & 0.0045 & \textbf{0.0052*} & 0.0715 & \textbf{0.1020*†} & \textbf{0.5662*} & 0.52464† \\ \hline
\end{tabular}
}

\subfloat[NDCG@10]{
\begin{tabular}{|cl|ll|ll|ll|ll|}
\hline
\multicolumn{2}{|l|}{} & \multicolumn{2}{l|}{MovieLens-20M} & \multicolumn{2}{l|}{Yelp} & \multicolumn{2}{l|}{Gowalla} & \multicolumn{2}{l|}{Booking.com} \\ \hline
\multicolumn{1}{|l|}{Architecture} & Loss & Cont & RSS & Cont & RSS & Cont & RSS & Cont & RSS \\ \hline
\multicolumn{1}{|c|}{\multirow{2}{*}{GRU4Rec}} & BCE & 0.0115† & \textbf{0.0183*} & 0.0035† & \textbf{0.0049*†} & \textbf{0.0017*} & 0.0002 & 0.2829 & \textbf{0.2899*} \\
\multicolumn{1}{|c|}{} & $\lambda$Rank & 0.0040 & \textbf{0.0839*†} & 0.0004 & \textbf{0.0014*} & 0.0033† & \textbf{0.0067*†} & \textbf{0.3132*†} & \textbf{0.3093†} \\ \hline
\multicolumn{1}{|c|}{\multirow{2}{*}{Caser}} & BCE & 0.0784† & \textbf{0.0995*} & 0.0021† & \textbf{0.0049*†} & 0.0039 & \textbf{0.0040} & \textbf{0.3665*†} & 0.3311† \\
\multicolumn{1}{|c|}{} & $\lambda$Rank & 0.0177 & \textbf{0.0814*†} & 0.0003 & \textbf{0.0007*} & 0.0055† & \textbf{0.0100*†} & 0.3181 & \textbf{0.3226*} \\ \hline
\multicolumn{1}{|c|}{\multirow{2}{*}{SASRec}} & BCE & 0.0850† & \textbf{0.1002*} & 0.0076† & \textbf{0.0136*†} & 0.0044 & \textbf{0.0044} & \textbf{0.3633*†} & 0.2966 \\
\multicolumn{1}{|c|}{} & $\lambda$Rank & 0.0579 & \textbf{0.1073*†} & 0.0021 & \textbf{0.0025*} & 0.0478† & \textbf{0.0749*†} & \textbf{0.3623*} & 0.3122† \\ \hline
\end{tabular}
}

\begin{tablenotes}
\item \footnotesize\textbf{Bold} denotes a more effective training objective for an (Architecture, Loss, Dataset) triplet. We use * to denote statistically significant differences compared to the other training objective (left vs.\ right), and $\dagger$ to denote significant differences on the change of loss function (upper vs.\ lower). All tests apply a paired t-test with Bonferroni multiple testing correction ($pvalue < 0.05$). \srs{Training time of all models is limited to 1 hour.}
\end{tablenotes}
\end{threeparttable}}
\end{table*}

\begin{table*}
\caption{Comparing RSS-enhanced SASRec with baseline models  under limited training.} \label{table:rssvsbaselines}
\begin{threeparttable}
\begin{tabular}{|ll|ll|ll|ll|ll|}
\hline
 &  & \multicolumn{2}{l|}{MovieLens-20M} & \multicolumn{2}{l|}{Yelp} & \multicolumn{2}{l|}{Gowalla} & \multicolumn{2}{l|}{Booking.com} \\ \hline
\multicolumn{1}{|l|}{Model} & \begin{tabular}[c]{@{}l@{}}Train\\ time\end{tabular} & \begin{tabular}[c]{@{}l@{}}Recall\\ @10\end{tabular} & \begin{tabular}[c]{@{}l@{}}NDCG\\ @10\end{tabular} & \begin{tabular}[c]{@{}l@{}}Recall\\ @10\end{tabular} & \begin{tabular}[c]{@{}l@{}}NDCG\\ @10\end{tabular} & \begin{tabular}[c]{@{}l@{}}Recall\\ @10\end{tabular} & \begin{tabular}[c]{@{}l@{}}NDCG\\ @10\end{tabular} & \begin{tabular}[c]{@{}l@{}}Recall\\ @10\end{tabular} & \begin{tabular}[c]{@{}l@{}}NDCG\\ @10\end{tabular} \\ \hline
\multicolumn{1}{|l|}{Popularity} & 1h & 0.049†* & 0.025†* & 0.006† & 0.003†* & 0.008* & 0.004* & 0.097†* & 0.043†* \\
\multicolumn{1}{|l|}{MF-BPR} & 1h & 0.079†* & 0.040†* & 0.019†* & 0.009†* & {\ul 0.029†*} & {\ul 0.018†*} & 0.449†* & 0.279†* \\
\multicolumn{1}{|l|}{SASRec-vanilla} & 1h & 0.136†* & 0.067†* & {\ul 0.022†*} & {\ul 0.011†*} & 0.010* & 0.005†* & 0.463†* & 0.270†* \\
\multicolumn{1}{|l|}{BERT4rec-1h} & 1h & 0.107†* & 0.053†* & 0.014†* & 0.007†* & N/A\textsuperscript{1} & N/A\textsuperscript{1} & 0.479†* & 0.288†* \\
\multicolumn{1}{|l|}{SASRec-RSS-BCE} & 1h & {\ul 0.189*} & {\ul 0.100*} & \textbf{0.027*} & \textbf{0.014*} & 0.009* & 0.004* & {\ul 0.518*} & {\ul 0.297*} \\
    \multicolumn{1}{|l|}{SASRec-RSS-$\lambda$Rank} & 1h & \textbf{0.197†} & \textbf{0.107†} & 0.005† & 0.003† & \textbf{0.102†} & \textbf{0.075†} & \textbf{0.525†} & \textbf{0.312†} \\ \hline
\multicolumn{1}{|l|}{BERT4Rec-16h\textsuperscript{2}} & 16h & 0.173†* & 0.092†* & 0.028* & 0.014* & N/A\textsuperscript{1} & N/A\textsuperscript{1} & 0.565†* & 0.354†* \\ \hline
\end{tabular}
\begin{tablenotes}
\item \footnotesize
\textbf{Bold} denotes the best model for a dataset by the metric in the main group, 
\underline{underlined} the second best. Symbols * and † denote statistically significant difference compared with SASRec-RSS-BCE and SASRec-RSS-$\lambda$Rank respectively, according to a paired t-test with Bonferroni multiple testing correction~($pvalue < 0.05$). 
\\
\textsuperscript{1} We do not report results for \bertrec{} models for the Gowalla dataset because due to large number of \crs{items in this} dataset, we were not able to train the model.  ~~~\textsuperscript{2} We report results for BERT4rec-16h separately due to its larger training time. 
\end{tablenotes}
\end{threeparttable}
\end{table*}

\subsection{Data Splitting and Evaluation Measures}\label{ssec:expsetup:eval}
Following many existing publications~\cite{sun2019bert4rec, kang2018sasrec, tang2018caser} we evaluate our method using a Leave-One-Out strategy. \srs{Specifically, for each user from  we hold out the final interaction \sasha{as} the test set, \srs{which we use to report metrics}. We also construct a validation set using the same Leave-One-Out strategy, using the second last interaction for a group of 1024 users as validation}. We set the number of training epochs to maximise NDCG@10 on the validation set.  For \craig{training}, we use \srs{all interactions except those included in the test and validation} \crs{sets}.

\hyphenation{Norma-lized}
We report two \craig{ranking evaluation measures}: Recall\footnote{In the context of sequential recommender systems, Recall corresponds to chances of correctly \crs{retrieving a single relevant item}, and therefore many publications~\cite{kang2018sasrec, sun2019bert4rec, yuan2019nextitnet}, call it Hit Ratio (HR). We prefer \craig{the} more \craig{conventional Recall} name for this metric.} and Normalized Discounted Cumulative Gain (NDCG). For both metrics, we apply \craig{a rank cutoff of 10}. To measure the significance of performances differences, we apply the paired t-test, and apply \sasha{Bonferroni} multiple testing correction, following recommended practices in IR~\cite{fuhr2021proof}.%

Until recently, for efficiency reasons, most of the sequential recommendations papers reported {\em sampled} metrics - i.e.\ they \craig{sampled} a small proportion of negative items and used only these items and the positive item when calculating evaluation measures. However, recent work by Krichene \& Rendle~\cite{rendle2020sampling} as well as Ca{\~n}amares \& Castells~\cite{canamares2020sampling} \crs{both showed} that using sampled metrics frequently leads to incorrect performance estimates, and the relative order of evaluated models can change. Hence in our experiments, we use full {\em unsampled} metrics: we rank all possible items in the catalog and calculate metrics on this full ranking\footnote{\craig{Indeed, we also found that conclusions could change using sampled metrics.}}.

\section{Results}\label{section:results}
\craig{We now analyse our experimental results for each of the \sasha{four} research questions stated in Section~\ref{ssec:expsetup:rq}.}

\hyphenation{archi-tec-tu-res}
\hyphenation{diffe-ren-ces}
\subsection{\ref{rq:improves}. Benefit of Recency Sampling}
To address our first research question, we compare our experimental architectures (GRU4Rec, Caser, SASRec) trained with either sequence continuation or RSS objectives. Table~\ref{table:contvsrss} reports the effectiveness results, in terms of Recall@10 and NDCG@10, of the three architectures, trained with both sequence continuation (denoted Cont) or RSS, and applying two different loss functions (Binary Cross-Entropy -- BCE -- and $\lambda$Rank)  on four datasets (MovieLens-20M, Yelp, Gowalla, Booking.com). Statistically significant differences -- according to a paired t-test with Bonferroni multiple testing correction ($pvalue < 0.05$) -- among the training objectives for a given architecture, model and loss function are shown.

On first inspection of Table~\ref{table:contvsrss}, we note that general magnitudes of the reported effectiveness results \sm{are smaller than those reported in~\cite{sun2019bert4rec}} - indeed, as stated in Section~\ref{ssec:expsetup:eval}, \sm{in contrast to~\cite{sun2019bert4rec}}, we follow recent advice~\cite{rendle2020sampling, canamares2020sampling} to avoid sampled metrics, instead preferring the more accurate unsampled metrics. The magnitudes of effectiveness reported for MovieLens-20M are in line with those reported by~\cite{dallmann2021case} (e.g.\ a  Recall@10 of 0.137 for SASRec-vanilla is reported in~\cite{dallmann2021case} when also using a Leave-One-Out evaluation scheme \sm{and unsampled metrics}).

We now turn to the comparison of training objectives. In particular, we note from the table that, on the MovieLens-20M, Yelp and Gowalla datasets, RSS results in improved NDCG@10 in 17 out of 18 cases -- 15 of which are by a statistically significant margin -- and also improved Recall@10 in 16 out of 18 cases (15 statistically significant). For instance, on MovieLens-20M, SASRec is the strongest performing architecture (in line with previous findings~\cite{kang2018sasrec, sun2019bert4rec}), however, applying RSS significantly improves its Recall@10, both when using BCE (0.153$\rightarrow$0.188) and when using $\lambda$Rank (0.105$\rightarrow$0.196). Similarly, and interestingly, SASRec with the RSS objective and  $\lambda$Rank loss outperformed other models by a very large margin on the Gowalla dataset (e.g. NDCG@10 0.102 vs.\ 0.071 when using sequence continuation). We postulate that the large number of items in the dataset made the training task very hard, and only the combination of RSS with $\lambda$Rank allows to train the model with reasonable quality in the given time limit -- this can be investigated further in future work. 

\sasha{On the other hand, for the Booking.com dataset, we observe that in 3 out of 6 cases, RSS is less effective. This is not an unexpected result: as we argued in Section~\ref{ssec:expsetup:datasets}, this dataset violates the underlying assumption encoded in Principle~\ref{principle:each_item}. Indeed, due to the geographical distance between items in this multi-city trip dataset, items cannot be considered out-of-order, and hence RSS does not improve the stronger models on this dataset.}

Overall, in response to RQ1, we conclude that Recency-based Sampling of Sequences improves the training of models if the items earlier in the user sequence can be treated as positives (propert\-ies exhibited by the MovieLens-20M and Gowalla datasets).

\subsection{\ref{rq:loss}. \srs{Comparison of Different Loss Functions}}

Next, we address the choice of loss function, as per \ref{rq:improves}. We again turn to Table~\ref{table:contvsrss}, but make comparisons of the upper vs.\ lower performances in each group. For instance, for RSS, we observe that applying the listwise $\lambda$Rank loss function on the GRU4Rec architecture on MovieLens-20M dataset results in a significant increase (0.035$\rightarrow$0.154), as denoted by the $\dagger$ symbol. Indeed, across all of Table~\ref{table:contvsrss}, we observe that when used with RSS training task, $\lambda$Rank improves NDCG@10 in 8 cases out of 12 (all 8 significantly) as well ass  Recall@10 in 8 cases out 12 (8 significantly). In contrast, $\lambda$Rank only improves over BCE in 7 out of 24 cases for the sequence continuation training objective (all by a significant margin). Overall, and in answer to RQ2, we find that $\lambda$Rank usually improves (\srs{except Yelp}) the effectiveness of our proposed RSS training objective, while it does not offer the same level of improvement for sequence continuation. We explain this finding as follows: in sequence continuation, there is only one relevant item per sequence, and hence the benefit of a listwise loss function is limited. In contrast, RSS selects multiple relevant items for each sequence, and in this case a listwise loss function can benefit in training the model to rank these items nearer the top of the ranking. However, as $\lambda$Rank did not improve RSS results on Yelp, we can not say that the improvements are consistent, and the question of the loss function selection requires further research.

\begin{figure*}
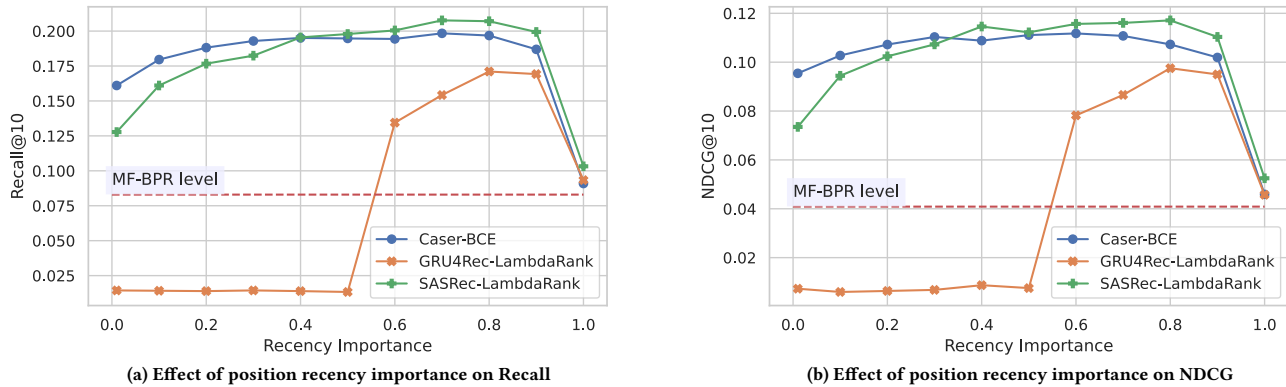

    \subfloat[Effect of position recency importance on Recall]{
        \label{subfig:importance_recall}
        \includesvg[width=0.5\linewidth]{recency_importance_recall.svg}
    }
    \subfloat[Effect of position recency importance on NDCG]{
        \label{subfig:importance_ndcg}
        \includesvg[width=0.5\linewidth]{recency_importance_ndcg.svg}
    }
    
  \caption{SASRec, GRU4rec and Caser performance on the MovieLens-20M dataset, when trained with Recency-based Sampling of Sequences with the exponential importance function $f(k) = \alpha^{(n-k)}$, where $n$ is the sequence length. Position recency importance parameter $\alpha$ is plotted on the $x$-axis. When $\alpha=0$,  the training objective turns into sequence continuation, and when $\alpha=1$ the task becomes similar to item masking or matrix reconstruction. Training time of all models is fixed at 1 hour. 
  }\label{fig:recency_importance}
\end{figure*}

\subsection{\ref{rq:recency}. Impact of Position Recency Importance}
This research question is concerned with the importance of  sampling
recent items in training sequences. To address this question, we train every experimental architecture with the best performing loss from Table~\ref{table:contvsrss} on the MovieLens-20M dataset. We vary the recency importance parameter $\alpha$ in the exponential recency importance function (Equation~\eqref{eq:exponential_importance}), to investigate its effect on effectiveness. In particular, as $\alpha \rightarrow 0$, the training task turns to sequence continuation, while large $\alpha$ the training task loses its sequential nature and becomes similar to matrix factorization.

Figure~\ref{fig:recency_importance} summarizes the impact of $\alpha$ on the \sm{model effectiveness}. We also present the performance of the MF-BPR~\cite{rendle2009bpr} baseline. From the figures, we observe that when we set $\alpha$ close to zero, the results match to those we report in Table~\ref{table:contvsrss} for the sequence continuation task, illustrating that under small $\alpha$, RSS only samples the last element in each sequence. Similarly, for $\alpha=1$ we observe that the effectiveness of all models drops almost to that of matrix factorization baseline, as target items are simply sampled from sequences without any ordering preference. \srs{Note, that in this case we sample target items uniformly, which is similar to BERT4Rec's item masking. However, BERT4Rec also has access to the positions of masked items \crs{(through the position embeddings)}, whereas in the case of $\alpha=1$ the positional information is completely lost and the model can not learn predicting the \emph{next} item and predicts \emph{some} item instead. Overall}, the general trends visible in Figure~\ref{fig:recency_importance} suggest that RSS allows to train effective models \crs{across} a wide range of the $\alpha$ parameter values: for Caser and SASRec, large improvements over sequence continuation training is achieved for $0.2 \leq \alpha \leq 0.9$; for GRU4Rec, strong performance is obtained $0.6 \leq \alpha \leq 0.9$. Indeed, for small $\alpha$, the number of positive items are limited, and hence the lambda gradients in $\lambda$Rank are also small. This provides little evidence to the GRUs in GRU4Rec, which therefore struggles with the vanishing gradient problem (a problem faced by many such recurrent architectures).

Overall, in response to RQ3, we find that the higher values of the recency importance parameter $\alpha\leq0.9$ result in effective performance for all three model architectures.

\subsection{\ref{rq:sota}. Comparison with Baselines} \label{ssec:sota}
To address our final research question concerning the comparison with baselines models, we compare the best performing RSS-enhanced model, SASRec-RSS, using both $\lambda$Rank and BCE losses, with the 5 baseline models described in Section~\ref{ssec:expsetup:models}.  Table~\ref{table:rssvsbaselines} summarizes the results of this comparison, reporting effectiveness metrics as well as training time durations. In particular, recall that all models are trained for less than 1 hour, except for BERT4Rec-16h (a full training of  BERT4Rec). Moreover, we did not train \bertrec{} on the Gowalla dataset, \cms{because the preprocessing code for \bertrec{} does not scale to its large number of items (indeed, Gowalla has more items than users, see Table~\ref{tab:datasets}). \sm{Indeed}, the preprocessing code to generate masked training sequences requires 14GB of storage for MovieLens-20M, but 548GB for Gowalla.}

On analysing Table~\ref{table:rssvsbaselines}, we observe that SASRec-RSS \srs{($\lambda$Rank or BCE)} achieves the most effective performance on all four datasets among the time-limited recommendation models. For instance, on the MovieLens-20M dataset, compared to the original formulation of SASRec (denoted SASRec-vanilla), the RSS adaptation significantly improves NDCG@10 (by a margin of 60\%) for the same training duration. Moreover, compared to the 16 hour training of \bertrec{}, SASRec-RSS exhibits 16\% higher NDCG@10 (a significant improvement), despite needing only 6\% of the training time (16h$\rightarrow$ 1h). For Booking.com, where RSS was less effective, SASRec-RSS with $\lambda$Rank objective obtains an NDCG@10 12\% less than that obtained by the expensive BER4Rec-16h model, and a Recall that is 7\% less. \srs{Interestingly, on the Yelp dataset, \crs{the} $\lambda$Rank version of SASRec is not effective (same performance as popularity baseline), but the BCE version of the model statistically significantly outperforms all other models in the main group and achieves performance on par with BERT4Rec-16h.} Furthermore, we see that \srs{in all cases where we able to train \bertrec{}}, under limited training time, \bertrec{} underperforms compared to the SASRec-RSS ($\lambda$Rank or BCE version).

Overall, in answer to RQ4, we find that SASRec-RSS can achieve significantly higher effectiveness than the state-of-the-art SASRec and \bertrec{} approaches when trained for a comparable time. Furthermore,  we can achieve performances exceeding or very close to a fully-trained \bertrec{}, but with much less training time. This highlights the importance of an appropriate training objective in general, and the benefits of our proposed RSS training objective in particular.

\section{Conclusions}\label{sec:conc}
In this work, we identified two limitations in existing training objectives for sequential recommender models. To address these two limitations, we proposed a refined training objective, called Recency-based Sampling of Sequences. Through experimentation on four datasets, we found that this relatively simple change in training objective can bring significant improvements in the overall effectiveness of state-of-the-art sequential recommendation models, such as SASRec and Caser. Furthermore, we showed that the $\lambda$Rank loss function brought further effectiveness benefits to training under RSS not otherwise observed under a more traditional sequence continuation task. Indeed, on the large MovieLens-20m dataset, we observed that RSS applied to the SASRec model can result in an \sasha{60\%} improvement in NDCG over the vanilla SASRec model, and a \sasha{16\%} improvement over a fully-trained BERT4Rec model, despite taking \sasha{93\%} less training time than BERT4Rec (see also Figure~\ref{fig:motivation}). Moreover, on the Yelp and Gowalla datasets, which both have geographic and strong sequential characteristics, RSS applied to SASRec brought significant benefits. Finally, while we did not apply RSS to \bertrec{}, due to its requirement of position embeddings (which it inherits from BERT), we believe that \bertrec{} could be adapted in future work to benefit from RSS.

\FloatBarrier

\bibliographystyle{ACM-Reference-Format}
\balance
\bibliography{references}


\begin{thebibliography}{55}


\ifx \showCODEN    \undefined \def \showCODEN     #1{\unskip}     \fi
\ifx \showDOI      \undefined \def \showDOI       #1{#1}\fi
\ifx \showISBNx    \undefined \def \showISBNx     #1{\unskip}     \fi
\ifx \showISBNxiii \undefined \def \showISBNxiii  #1{\unskip}     \fi
\ifx \showISSN     \undefined \def \showISSN      #1{\unskip}     \fi
\ifx \showLCCN     \undefined \def \showLCCN      #1{\unskip}     \fi
\ifx \shownote     \undefined \def \shownote      #1{#1}          \fi
\ifx \showarticletitle \undefined \def \showarticletitle #1{#1}   \fi
\ifx \showURL      \undefined \def \showURL       {\relax}        \fi
\providecommand\bibfield[2]{#2}
\providecommand\bibinfo[2]{#2}
\providecommand\natexlab[1]{#1}
\providecommand\showeprint[2][]{arXiv:#2}

\bibitem[Abadi et~al\mbox{.}(2016)]%
        {abadi2016tensorflow}
\bibfield{author}{\bibinfo{person}{Mart{\'\i}n Abadi}, \bibinfo{person}{Paul
  Barham}, \bibinfo{person}{Jianmin Chen}, \bibinfo{person}{Zhifeng Chen},
  \bibinfo{person}{Andy Davis}, \bibinfo{person}{Jeffrey Dean},
  \bibinfo{person}{Matthieu Devin}, \bibinfo{person}{Sanjay Ghemawat},
  \bibinfo{person}{Geoffrey Irving}, \bibinfo{person}{Michael Isard},
  {et~al\mbox{.}}} \bibinfo{year}{2016}\natexlab{}.
\newblock \showarticletitle{{TensorFlow}: A system for large-scale machine
  learning}. In \bibinfo{booktitle}{\emph{Proc. USENIX}}.
  \bibinfo{pages}{265--283}.
\newblock


\bibitem[Amjadi et~al\mbox{.}(2021)]%
        {amjadi2021katrec}
\bibfield{author}{\bibinfo{person}{Mehrnaz Amjadi},
  \bibinfo{person}{Seyed~Danial Mohseni~Taheri}, {and} \bibinfo{person}{Theja
  Tulabandhula}.} \bibinfo{year}{2021}\natexlab{}.
\newblock \showarticletitle{{KATRec}: Knowledge aware attentive sequential
  recommendations}. In \bibinfo{booktitle}{\emph{Proc. ICDS}}.
  \bibinfo{pages}{305--320}.
\newblock


\bibitem[Bian et~al\mbox{.}(2021)]%
        {bian2021contrastive}
\bibfield{author}{\bibinfo{person}{Shuqing Bian}, \bibinfo{person}{Wayne~Xin
  Zhao}, \bibinfo{person}{Kun Zhou}, \bibinfo{person}{Jing Cai},
  \bibinfo{person}{Yancheng He}, \bibinfo{person}{Cunxiang Yin}, {and}
  \bibinfo{person}{Ji-Rong Wen}.} \bibinfo{year}{2021}\natexlab{}.
\newblock \showarticletitle{Contrastive Curriculum Learning for Sequential User
  Behavior Modeling via Data Augmentation}. In \bibinfo{booktitle}{\emph{Proc.
  CIKM}}. \bibinfo{pages}{3737--3746}.
\newblock


\bibitem[Burges(2010)]%
        {burges2010ranknet}
\bibfield{author}{\bibinfo{person}{Christopher~JC Burges}.}
  \bibinfo{year}{2010}\natexlab{}.
\newblock \showarticletitle{From {RankNet} to {LambdaRank} to {LambdaMART}: An
  overview}.
\newblock \bibinfo{journal}{\emph{Learning}} \bibinfo{volume}{11},
  \bibinfo{number}{23-581} (\bibinfo{year}{2010}), \bibinfo{pages}{81}.
\newblock


\bibitem[Ca{\~n}amares and Castells(2020)]%
        {canamares2020sampling}
\bibfield{author}{\bibinfo{person}{Roc{\'\i}o Ca{\~n}amares} {and}
  \bibinfo{person}{Pablo Castells}.} \bibinfo{year}{2020}\natexlab{}.
\newblock \showarticletitle{On target item sampling in offline recommender
  system evaluation}. In \bibinfo{booktitle}{\emph{Proc. RecSys}}.
  \bibinfo{pages}{259--268}.
\newblock


\bibitem[Chapelle and Chang(2011)]%
        {chapelle2011yahoo}
\bibfield{author}{\bibinfo{person}{Olivier Chapelle} {and} \bibinfo{person}{Yi
  Chang}.} \bibinfo{year}{2011}\natexlab{}.
\newblock \showarticletitle{Yahoo! learning to rank challenge overview}.
\newblock \bibinfo{journal}{\emph{Proceedings of Machine Learning Research}}
  (\bibinfo{year}{2011}), \bibinfo{pages}{1--24}.
\newblock


\bibitem[Cho et~al\mbox{.}(2011)]%
        {cho2011gowalla}
\bibfield{author}{\bibinfo{person}{Eunjoon Cho}, \bibinfo{person}{Seth~A
  Myers}, {and} \bibinfo{person}{Jure Leskovec}.}
  \bibinfo{year}{2011}\natexlab{}.
\newblock \showarticletitle{Friendship and mobility: user movement in
  location-based social networks}. In \bibinfo{booktitle}{\emph{Proc. KDD}}.
  \bibinfo{pages}{1082--1090}.
\newblock


\bibitem[Cho et~al\mbox{.}(2020)]%
        {cho2020meantime}
\bibfield{author}{\bibinfo{person}{Sung~Min Cho}, \bibinfo{person}{Eunhyeok
  Park}, {and} \bibinfo{person}{Sungjoo Yoo}.} \bibinfo{year}{2020}\natexlab{}.
\newblock \showarticletitle{MEANTIME: Mixture of attention mechanisms with
  multi-temporal embeddings for sequential recommendation}. In
  \bibinfo{booktitle}{\emph{Proc. RecSys}}. \bibinfo{pages}{515--520}.
\newblock


\bibitem[Dallmann et~al\mbox{.}(2021)]%
        {dallmann2021case}
\bibfield{author}{\bibinfo{person}{Alexander Dallmann}, \bibinfo{person}{Daniel
  Zoller}, {and} \bibinfo{person}{Andreas Hotho}.}
  \bibinfo{year}{2021}\natexlab{}.
\newblock \showarticletitle{A Case Study on Sampling Strategies for Evaluating
  Neural Sequential Item Recommendation Models}. In
  \bibinfo{booktitle}{\emph{Proc. RecSys}}. \bibinfo{pages}{505--514}.
\newblock


\bibitem[Devlin et~al\mbox{.}(2019)]%
        {devlin2018bert}
\bibfield{author}{\bibinfo{person}{Jacob Devlin}, \bibinfo{person}{Ming-Wei
  Chang}, \bibinfo{person}{Kenton Lee}, {and} \bibinfo{person}{Kristina
  Toutanova}.} \bibinfo{year}{2019}\natexlab{}.
\newblock \showarticletitle{{BERT}: Pre-training of Deep Bidirectional
  Transformers for Language Understanding}. In \bibinfo{booktitle}{\emph{Proc.
  NAACL-HLT}}. \bibinfo{pages}{4171--4186}.
\newblock


\bibitem[Fischer et~al\mbox{.}(2020)]%
        {fischer2020kebert}
\bibfield{author}{\bibinfo{person}{Elisabeth Fischer}, \bibinfo{person}{Daniel
  Zoller}, \bibinfo{person}{Alexander Dallmann}, {and} \bibinfo{person}{Andreas
  Hotho}.} \bibinfo{year}{2020}\natexlab{}.
\newblock \showarticletitle{Integrating keywords into {BERT4Rec} for sequential
  recommendation}. In \bibinfo{booktitle}{\emph{German Conference on Artificial
  Intelligence (K{\"u}nstliche Intelligenz)}}. \bibinfo{pages}{275--282}.
\newblock


\bibitem[Fuhr(2021)]%
        {fuhr2021proof}
\bibfield{author}{\bibinfo{person}{Norbert Fuhr}.}
  \bibinfo{year}{2021}\natexlab{}.
\newblock \showarticletitle{Proof by experimentation? Towards better {IR}
  research}. In \bibinfo{booktitle}{\emph{ACM SIGIR Forum}},
  Vol.~\bibinfo{volume}{54}. \bibinfo{pages}{1--4}.
\newblock


\bibitem[Goldenberg and Levin(2021)]%
        {goldenberg2021booking}
\bibfield{author}{\bibinfo{person}{Dmitri Goldenberg} {and}
  \bibinfo{person}{Pavel Levin}.} \bibinfo{year}{2021}\natexlab{}.
\newblock \showarticletitle{Booking.com Multi-Destination Trips Dataset}. In
  \bibinfo{booktitle}{\emph{Proc. SIGIR}}. \bibinfo{pages}{2457--2462}.
\newblock


\bibitem[Harper and Konstan(2015)]%
        {harper2015movielens}
\bibfield{author}{\bibinfo{person}{F~Maxwell Harper} {and}
  \bibinfo{person}{Joseph~A Konstan}.} \bibinfo{year}{2015}\natexlab{}.
\newblock \showarticletitle{The {MovieLens} datasets: History and context}.
\newblock \bibinfo{journal}{\emph{ACM Transactions on Interactive Intelligent
  Systems (TIIS)}} \bibinfo{volume}{5}, \bibinfo{number}{4}
  (\bibinfo{year}{2015}), \bibinfo{pages}{1--19}.
\newblock


\bibitem[Hidasi and Karatzoglou(2018)]%
        {hidasi2018gru4recplus}
\bibfield{author}{\bibinfo{person}{Bal{\'a}zs Hidasi} {and}
  \bibinfo{person}{Alexandros Karatzoglou}.} \bibinfo{year}{2018}\natexlab{}.
\newblock \showarticletitle{Recurrent neural networks with top-k gains for
  session-based recommendations}. In \bibinfo{booktitle}{\emph{Proc. CIKM}}.
  \bibinfo{pages}{843--852}.
\newblock


\bibitem[Hidasi et~al\mbox{.}(2016)]%
        {hidasi2015gru4rec}
\bibfield{author}{\bibinfo{person}{Bal{\'{a}}zs Hidasi},
  \bibinfo{person}{Alexandros Karatzoglou}, \bibinfo{person}{Linas Baltrunas},
  {and} \bibinfo{person}{Domonkos Tikk}.} \bibinfo{year}{2016}\natexlab{}.
\newblock \showarticletitle{Session-based Recommendations with Recurrent Neural
  Networks}. In \bibinfo{booktitle}{\emph{Proc. ICLR}}.
\newblock


\bibitem[Huang et~al\mbox{.}(2018)]%
        {huang2018improving}
\bibfield{author}{\bibinfo{person}{Jin Huang}, \bibinfo{person}{Wayne~Xin
  Zhao}, \bibinfo{person}{Hongjian Dou}, \bibinfo{person}{Ji-Rong Wen}, {and}
  \bibinfo{person}{Edward~Y Chang}.} \bibinfo{year}{2018}\natexlab{}.
\newblock \showarticletitle{Improving sequential recommendation with
  knowledge-enhanced memory networks}. In \bibinfo{booktitle}{\emph{Proc.
  SIGIR}}. \bibinfo{pages}{505--514}.
\newblock


\bibitem[Kang and McAuley(2018)]%
        {kang2018sasrec}
\bibfield{author}{\bibinfo{person}{Wang-Cheng Kang} {and}
  \bibinfo{person}{Julian McAuley}.} \bibinfo{year}{2018}\natexlab{}.
\newblock \showarticletitle{Self-attentive sequential recommendation}. In
  \bibinfo{booktitle}{\emph{Proc. ICDM}}. \bibinfo{pages}{197--206}.
\newblock


\bibitem[Koopmann et~al\mbox{.}(2021)]%
        {koopmann2021cobert}
\bibfield{author}{\bibinfo{person}{Tobias Koopmann},
  \bibinfo{person}{Konstantin Kobs}, \bibinfo{person}{Konstantin Herud}, {and}
  \bibinfo{person}{Andreas Hotho}.} \bibinfo{year}{2021}\natexlab{}.
\newblock \showarticletitle{{CoBERT}: Scientific Collaboration Prediction via
  Sequential Recommendation}. In \bibinfo{booktitle}{\emph{Proc. ICDMW}}.
  \bibinfo{pages}{45--54}.
\newblock


\bibitem[Krichene and Rendle(2020)]%
        {rendle2020sampling}
\bibfield{author}{\bibinfo{person}{Walid Krichene} {and}
  \bibinfo{person}{Steffen Rendle}.} \bibinfo{year}{2020}\natexlab{}.
\newblock \showarticletitle{On sampled metrics for item recommendation}. In
  \bibinfo{booktitle}{\emph{Proc. KDD}}. \bibinfo{pages}{1748--1757}.
\newblock


\bibitem[Kula(2015)]%
        {kula2015lightfm}
\bibfield{author}{\bibinfo{person}{Maciej Kula}.}
  \bibinfo{year}{2015}\natexlab{}.
\newblock \showarticletitle{Metadata Embeddings for User and Item Cold-start
  Recommendations}. In \bibinfo{booktitle}{\emph{Proc. Workshop on New Trends
  on Content-Based Recommender @ RecSys}} \emph{(\bibinfo{series}{{CEUR}
  Workshop Proc.}, Vol.~\bibinfo{volume}{1448})}. \bibinfo{pages}{14--21}.
\newblock


\bibitem[Lee et~al\mbox{.}(2021)]%
        {lee2021moi}
\bibfield{author}{\bibinfo{person}{Hojoon Lee}, \bibinfo{person}{Dongyoon
  Hwang}, \bibinfo{person}{Sunghwan Hong}, \bibinfo{person}{Changyeon Kim},
  \bibinfo{person}{Seungryong Kim}, {and} \bibinfo{person}{Jaegul Choo}.}
  \bibinfo{year}{2021}\natexlab{}.
\newblock \showarticletitle{{MOI-Mixer}: Improving {MLP-Mixer} with Multi Order
  Interactions in Sequential Recommendation}.
\newblock \bibinfo{journal}{\emph{arXiv preprint arXiv:2108.07505}}
  (\bibinfo{year}{2021}).
\newblock


\bibitem[Li et~al\mbox{.}(2021b)]%
        {li2021intention}
\bibfield{author}{\bibinfo{person}{Haoyang Li}, \bibinfo{person}{Xin Wang},
  \bibinfo{person}{Ziwei Zhang}, \bibinfo{person}{Jianxin Ma},
  \bibinfo{person}{Peng Cui}, {and} \bibinfo{person}{Wenwu Zhu}.}
  \bibinfo{year}{2021}\natexlab{b}.
\newblock \showarticletitle{Intention-aware sequential recommendation with
  structured intent transition}.
\newblock \bibinfo{journal}{\emph{IEEE Transactions on Knowledge and Data
  Engineering (TKDE)}} (\bibinfo{year}{2021}).
\newblock


\bibitem[Li et~al\mbox{.}(2021a)]%
        {li2021new}
\bibfield{author}{\bibinfo{person}{Roger~Zhe Li}, \bibinfo{person}{Juli{\'a}n
  Urbano}, {and} \bibinfo{person}{Alan Hanjalic}.}
  \bibinfo{year}{2021}\natexlab{a}.
\newblock \showarticletitle{New Insights into Metric Optimization for
  Ranking-based Recommendation}. In \bibinfo{booktitle}{\emph{Proc. SIGIR}}.
  \bibinfo{pages}{932–941}.
\newblock


\bibitem[Liu(2009)]%
        {Liu09ftir}
\bibfield{author}{\bibinfo{person}{Tie-Yan Liu}.}
  \bibinfo{year}{2009}\natexlab{}.
\newblock \showarticletitle{Learning to Rank for Information Retrieval}.
\newblock \bibinfo{journal}{\emph{Foundations and Trends in Information
  Retrieval}} \bibinfo{volume}{3}, \bibinfo{number}{3} (\bibinfo{year}{2009}),
  \bibinfo{pages}{225--331}.
\newblock


\bibitem[Liu et~al\mbox{.}(2021)]%
        {liu2021augmenting}
\bibfield{author}{\bibinfo{person}{Zhiwei Liu}, \bibinfo{person}{Ziwei Fan},
  \bibinfo{person}{Yu Wang}, {and} \bibinfo{person}{Philip~S. Yu}.}
  \bibinfo{year}{2021}\natexlab{}.
\newblock \showarticletitle{Augmenting Sequential Recommendation with
  Pseudo-Prior Items via Reversely Pre-training Transformer}. In
  \bibinfo{booktitle}{\emph{Proc. SIGIR}}. \bibinfo{pages}{1608–1612}.
\newblock


\bibitem[Ma et~al\mbox{.}(2019)]%
        {ma2019hierarchical}
\bibfield{author}{\bibinfo{person}{Chen Ma}, \bibinfo{person}{Peng Kang}, {and}
  \bibinfo{person}{Xue Liu}.} \bibinfo{year}{2019}\natexlab{}.
\newblock \showarticletitle{Hierarchical gating networks for sequential
  recommendation}. In \bibinfo{booktitle}{\emph{Proc. KDD}}.
  \bibinfo{pages}{825--833}.
\newblock


\bibitem[Ma et~al\mbox{.}(2020)]%
        {ma2020disentangled}
\bibfield{author}{\bibinfo{person}{Jianxin Ma}, \bibinfo{person}{Chang Zhou},
  \bibinfo{person}{Hongxia Yang}, \bibinfo{person}{Peng Cui},
  \bibinfo{person}{Xin Wang}, {and} \bibinfo{person}{Wenwu Zhu}.}
  \bibinfo{year}{2020}\natexlab{}.
\newblock \showarticletitle{Disentangled self-supervision in sequential
  recommenders}. In \bibinfo{booktitle}{\emph{Proc. KDD}}.
  \bibinfo{pages}{483--491}.
\newblock


\bibitem[Meng et~al\mbox{.}(2021)]%
        {meng2021vbcar}
\bibfield{author}{\bibinfo{person}{Zaiqiao Meng}, \bibinfo{person}{Richard
  McCreadie}, \bibinfo{person}{Craig Macdonald}, {and} \bibinfo{person}{Iadh
  Ounis}.} \bibinfo{year}{2021}\natexlab{}.
\newblock \showarticletitle{Variational Bayesian representation learning for
  grocery recommendation}.
\newblock \bibinfo{journal}{\emph{Information Retrieval Journal}}
  \bibinfo{volume}{24} (\bibinfo{date}{10} \bibinfo{year}{2021}),
  \bibinfo{pages}{1--23}.
\newblock


\bibitem[Padungkiatwattana et~al\mbox{.}(2022)]%
        {padungkiatwattana2022arerec}
\bibfield{author}{\bibinfo{person}{Umaporn Padungkiatwattana},
  \bibinfo{person}{Thitiya Sae-Diae}, \bibinfo{person}{Saranya Maneeroj}, {and}
  \bibinfo{person}{Atsuhiro Takasu}.} \bibinfo{year}{2022}\natexlab{}.
\newblock \showarticletitle{ARERec: Attentive Local Interaction Model for
  Sequential Recommendation}.
\newblock \bibinfo{journal}{\emph{IEEE Access}}.
\newblock


\bibitem[Petrov and Macdonald(2022)]%
        {petrov2022replicability}
\bibfield{author}{\bibinfo{person}{Aleksandr Petrov} {and}
  \bibinfo{person}{Craig Macdonald}.} \bibinfo{year}{2022}\natexlab{}.
\newblock \showarticletitle{A Systematic Review and Replicability Study of
  BERT4Rec for Sequential Recommendation}. In \bibinfo{booktitle}{\emph{Proc.
  RecSys}}.
\newblock


\bibitem[Petrov and Makarov(2021)]%
        {petrov2021booking}
\bibfield{author}{\bibinfo{person}{Aleksandr Petrov} {and}
  \bibinfo{person}{Yuriy Makarov}.} \bibinfo{year}{2021}\natexlab{}.
\newblock \showarticletitle{Attention-based neural re-ranking approach for next
  city in trip recommendations}. In \bibinfo{booktitle}{\emph{Proc. WSDM
  WebTour}}. \bibinfo{pages}{41--45}.
\newblock


\bibitem[Qin et~al\mbox{.}(2021)]%
        {qin2020neural}
\bibfield{author}{\bibinfo{person}{Zhen Qin}, \bibinfo{person}{Le Yan},
  \bibinfo{person}{Honglei Zhuang}, \bibinfo{person}{Yi Tay},
  \bibinfo{person}{Rama~Kumar Pasumarthi}, \bibinfo{person}{Xuanhui Wang},
  \bibinfo{person}{Michael Bendersky}, {and} \bibinfo{person}{Marc Najork}.}
  \bibinfo{year}{2021}\natexlab{}.
\newblock \showarticletitle{Are Neural Rankers still Outperformed by Gradient
  Boosted Decision Trees?}. In \bibinfo{booktitle}{\emph{Proc. ICLR}}.
\newblock


\bibitem[Qiu et~al\mbox{.}(2021b)]%
        {qiu2021GraphExploiting}
\bibfield{author}{\bibinfo{person}{Ruihong Qiu}, \bibinfo{person}{Zi Huang},
  \bibinfo{person}{Tong Chen}, {and} \bibinfo{person}{Hongzhi Yin}.}
  \bibinfo{year}{2021}\natexlab{b}.
\newblock \showarticletitle{Exploiting Positional Information for Session-based
  Recommendation}.
\newblock \bibinfo{journal}{\emph{ACM Transactions on Information Systems
  (TOIS)}} \bibinfo{volume}{40}, \bibinfo{number}{2} (\bibinfo{year}{2021}),
  \bibinfo{pages}{1--24}.
\newblock


\bibitem[Qiu et~al\mbox{.}(2020a)]%
        {qiu2020GraphExploiting}
\bibfield{author}{\bibinfo{person}{Ruihong Qiu}, \bibinfo{person}{Zi Huang},
  \bibinfo{person}{Jingjing Li}, {and} \bibinfo{person}{Hongzhi Yin}.}
  \bibinfo{year}{2020}\natexlab{a}.
\newblock \showarticletitle{Exploiting cross-session information for
  session-based recommendation with graph neural networks}.
\newblock \bibinfo{journal}{\emph{ACM Transactions on Information Systems
  (TOIS)}} \bibinfo{volume}{38}, \bibinfo{number}{3} (\bibinfo{year}{2020}),
  \bibinfo{pages}{1--23}.
\newblock


\bibitem[Qiu et~al\mbox{.}(2021a)]%
        {qiu2021GraphMemory}
\bibfield{author}{\bibinfo{person}{Ruihong Qiu}, \bibinfo{person}{Zi Huang},
  {and} \bibinfo{person}{Hongzhi Yin}.} \bibinfo{year}{2021}\natexlab{a}.
\newblock \showarticletitle{Memory Augmented Multi-Instance Contrastive
  Predictive Coding for Sequential Recommendation}.
\newblock \bibinfo{journal}{\emph{CoRR}}  \bibinfo{volume}{abs/2109.00368}
  (\bibinfo{year}{2021}).
\newblock


\bibitem[Qiu et~al\mbox{.}(2022)]%
        {qiu2022contrastive}
\bibfield{author}{\bibinfo{person}{Ruihong Qiu}, \bibinfo{person}{Zi Huang},
  \bibinfo{person}{Hongzhi Yin}, {and} \bibinfo{person}{Zijian Wang}.}
  \bibinfo{year}{2022}\natexlab{}.
\newblock \showarticletitle{Contrastive learning for representation
  degeneration problem in sequential recommendation}. In
  \bibinfo{booktitle}{\emph{Proc. WSDM}}. \bibinfo{pages}{813--823}.
\newblock


\bibitem[Qiu et~al\mbox{.}(2020b)]%
        {Graph2020gag}
\bibfield{author}{\bibinfo{person}{Ruihong Qiu}, \bibinfo{person}{Hongzhi Yin},
  \bibinfo{person}{Zi Huang}, {and} \bibinfo{person}{Tong Chen}.}
  \bibinfo{year}{2020}\natexlab{b}.
\newblock \showarticletitle{{GAG}: Global attributed graph neural network for
  streaming session-based recommendation}. In \bibinfo{booktitle}{\emph{Proc.
  SIGIR}}. \bibinfo{pages}{669--678}.
\newblock


\bibitem[Quadrana et~al\mbox{.}(2018)]%
        {quadrana2018sequence}
\bibfield{author}{\bibinfo{person}{Massimo Quadrana}, \bibinfo{person}{Paolo
  Cremonesi}, {and} \bibinfo{person}{Dietmar Jannach}.}
  \bibinfo{year}{2018}\natexlab{}.
\newblock \showarticletitle{Sequence-aware recommender systems}.
\newblock \bibinfo{journal}{\emph{ACM Computing Surveys (CSUR)}}
  \bibinfo{volume}{51}, \bibinfo{number}{4} (\bibinfo{year}{2018}),
  \bibinfo{pages}{1--36}.
\newblock


\bibitem[Rendle et~al\mbox{.}(2009)]%
        {rendle2009bpr}
\bibfield{author}{\bibinfo{person}{Steffen Rendle}, \bibinfo{person}{Christoph
  Freudenthaler}, \bibinfo{person}{Zeno Gantner}, {and} \bibinfo{person}{Lars
  Schmidt-Thieme}.} \bibinfo{year}{2009}\natexlab{}.
\newblock \showarticletitle{{BPR}: Bayesian personalized ranking from implicit
  feedback}. In \bibinfo{booktitle}{\emph{Proc. CUAI}}.
  \bibinfo{pages}{452--461}.
\newblock


\bibitem[Rendle et~al\mbox{.}(2010)]%
        {rendle2010fpmc}
\bibfield{author}{\bibinfo{person}{Steffen Rendle}, \bibinfo{person}{Christoph
  Freudenthaler}, {and} \bibinfo{person}{Lars Schmidt-Thieme}.}
  \bibinfo{year}{2010}\natexlab{}.
\newblock \showarticletitle{Factorizing personalized markov chains for
  next-basket recommendation}. In \bibinfo{booktitle}{\emph{Proc. WWW}}.
  \bibinfo{pages}{811--820}.
\newblock


\bibitem[Sun et~al\mbox{.}(2019)]%
        {sun2019bert4rec}
\bibfield{author}{\bibinfo{person}{Fei Sun}, \bibinfo{person}{Jun Liu},
  \bibinfo{person}{Jian Wu}, \bibinfo{person}{Changhua Pei},
  \bibinfo{person}{Xiao Lin}, \bibinfo{person}{Wenwu Ou}, {and}
  \bibinfo{person}{Peng Jiang}.} \bibinfo{year}{2019}\natexlab{}.
\newblock \showarticletitle{{BERT4Rec}: Sequential recommendation with
  bidirectional encoder representations from transformer}. In
  \bibinfo{booktitle}{\emph{Proc. CIKM}}. \bibinfo{pages}{1441--1450}.
\newblock


\bibitem[Tang and Wang(2018)]%
        {tang2018caser}
\bibfield{author}{\bibinfo{person}{Jiaxi Tang} {and} \bibinfo{person}{Ke
  Wang}.} \bibinfo{year}{2018}\natexlab{}.
\newblock \showarticletitle{Personalized top-n sequential recommendation via
  convolutional sequence embedding}. In \bibinfo{booktitle}{\emph{Proc. WSDM}}.
  \bibinfo{pages}{565--573}.
\newblock


\bibitem[Vaswani et~al\mbox{.}(2017)]%
        {vaswani2017attention}
\bibfield{author}{\bibinfo{person}{Ashish Vaswani}, \bibinfo{person}{Noam
  Shazeer}, \bibinfo{person}{Niki Parmar}, \bibinfo{person}{Jakob Uszkoreit},
  \bibinfo{person}{Llion Jones}, \bibinfo{person}{Aidan~N Gomez},
  \bibinfo{person}{{\L}ukasz Kaiser}, {and} \bibinfo{person}{Illia
  Polosukhin}.} \bibinfo{year}{2017}\natexlab{}.
\newblock \showarticletitle{Attention is all you need}. In
  \bibinfo{booktitle}{\emph{Proc. NeurIPS}}. \bibinfo{pages}{5998--6008}.
\newblock


\bibitem[Wan et~al\mbox{.}(2018)]%
        {wan2018representing}
\bibfield{author}{\bibinfo{person}{Mengting Wan}, \bibinfo{person}{Di Wang},
  \bibinfo{person}{Jie Liu}, \bibinfo{person}{Paul Bennett}, {and}
  \bibinfo{person}{Julian McAuley}.} \bibinfo{year}{2018}\natexlab{}.
\newblock \showarticletitle{Representing and recommending shopping baskets with
  complementarity, compatibility and loyalty}. In
  \bibinfo{booktitle}{\emph{Proc. CIKM}}. \bibinfo{pages}{1133--1142}.
\newblock


\bibitem[Wang et~al\mbox{.}(2022)]%
        {wang2022sequential}
\bibfield{author}{\bibinfo{person}{Chenyang Wang}, \bibinfo{person}{Weizhi Ma},
  {and} \bibinfo{person}{Chong Chen}.} \bibinfo{year}{2022}\natexlab{}.
\newblock \showarticletitle{Sequential Recommendation with Multiple Contrast
  Signals}.
\newblock \bibinfo{journal}{\emph{ACM Transactions on Information Systems
  (TOIS)}} (\bibinfo{year}{2022}).
\newblock


\bibitem[Wu et~al\mbox{.}(2021)]%
        {wu2021seq2bubbles}
\bibfield{author}{\bibinfo{person}{Qitian Wu}, \bibinfo{person}{Chenxiao Yang},
  \bibinfo{person}{Shuodian Yu}, \bibinfo{person}{Xiaofeng Gao}, {and}
  \bibinfo{person}{Guihai Chen}.} \bibinfo{year}{2021}\natexlab{}.
\newblock \showarticletitle{Seq2Bubbles: Region-Based Embedding Learning for
  User Behaviors in Sequential Recommenders}. In
  \bibinfo{booktitle}{\emph{Proc. CIKM}}. \bibinfo{pages}{2160--2169}.
\newblock


\bibitem[Xie et~al\mbox{.}(2020)]%
        {xie2020contrastive}
\bibfield{author}{\bibinfo{person}{Xu Xie}, \bibinfo{person}{Fei Sun},
  \bibinfo{person}{Zhaoyang Liu}, \bibinfo{person}{Shiwen Wu},
  \bibinfo{person}{Jinyang Gao}, \bibinfo{person}{Bolin Ding}, {and}
  \bibinfo{person}{Bin Cui}.} \bibinfo{year}{2020}\natexlab{}.
\newblock \showarticletitle{Contrastive Learning for Sequential
  Recommendation}.
\newblock \bibinfo{journal}{\emph{arXiv preprint arXiv:2010.14395}}
  (\bibinfo{year}{2020}).
\newblock


\bibitem[Yuan et~al\mbox{.}(2016)]%
        {yuan2016lambdafm}
\bibfield{author}{\bibinfo{person}{Fajie Yuan}, \bibinfo{person}{Guibing Guo},
  \bibinfo{person}{Joemon~M Jose}, \bibinfo{person}{Long Chen},
  \bibinfo{person}{Haitao Yu}, {and} \bibinfo{person}{Weinan Zhang}.}
  \bibinfo{year}{2016}\natexlab{}.
\newblock \showarticletitle{{LambdaFM}: learning optimal ranking with
  factorization machines using lambda surrogates}. In
  \bibinfo{booktitle}{\emph{Proc. CIKM}}. \bibinfo{pages}{227--236}.
\newblock


\bibitem[Yuan et~al\mbox{.}(2019)]%
        {yuan2019nextitnet}
\bibfield{author}{\bibinfo{person}{Fajie Yuan}, \bibinfo{person}{Alexandros
  Karatzoglou}, \bibinfo{person}{Ioannis Arapakis}, \bibinfo{person}{Joemon~M
  Jose}, {and} \bibinfo{person}{Xiangnan He}.} \bibinfo{year}{2019}\natexlab{}.
\newblock \showarticletitle{A simple convolutional generative network for next
  item recommendation}. In \bibinfo{booktitle}{\emph{Proc. WSDM}}.
  \bibinfo{pages}{582--590}.
\newblock


\bibitem[Zhan et~al\mbox{.}(2022)]%
        {zhan2022transrecplusplus}
\bibfield{author}{\bibinfo{person}{Zhuo-Xin Zhan}, \bibinfo{person}{Ming-Kai
  He}, \bibinfo{person}{Wei-Ke Pan}, {and} \bibinfo{person}{Zhong Ming}.}
  \bibinfo{year}{2022}\natexlab{}.
\newblock \showarticletitle{Transrec++: Translation-based sequential
  recommendation with heterogeneous feedback}.
\newblock \bibinfo{journal}{\emph{Frontiers of Computer Science}}
  \bibinfo{volume}{16}, \bibinfo{number}{2} (\bibinfo{year}{2022}),
  \bibinfo{pages}{1--3}.
\newblock


\bibitem[Zhang et~al\mbox{.}(2022)]%
        {zhang2022dynamic}
\bibfield{author}{\bibinfo{person}{Mengqi Zhang}, \bibinfo{person}{Shu Wu},
  \bibinfo{person}{Xueli Yu}, \bibinfo{person}{Qiang Liu}, {and}
  \bibinfo{person}{Liang Wang}.} \bibinfo{year}{2022}\natexlab{}.
\newblock \showarticletitle{Dynamic graph neural networks for sequential
  recommendation}.
\newblock \bibinfo{journal}{\emph{IEEE Transactions on Knowledge and Data
  Engineering (TKDE)}} (\bibinfo{year}{2022}).
\newblock


\bibitem[Zhao et~al\mbox{.}(2021)]%
        {zhao2021adversarial}
\bibfield{author}{\bibinfo{person}{Pengyu Zhao}, \bibinfo{person}{Tianxiao
  Shui}, \bibinfo{person}{Yuanxing Zhang}, \bibinfo{person}{Kecheng Xiao},
  {and} \bibinfo{person}{Kaigui Bian}.} \bibinfo{year}{2021}\natexlab{}.
\newblock \showarticletitle{Adversarial oracular seq2seq learning for
  sequential recommendation}. In \bibinfo{booktitle}{\emph{Proc. ICJAI}}.
\newblock


\bibitem[Zhou et~al\mbox{.}(2020)]%
        {zhou2020s3}
\bibfield{author}{\bibinfo{person}{Kun Zhou}, \bibinfo{person}{Hui Wang},
  \bibinfo{person}{Wayne~Xin Zhao}, \bibinfo{person}{Yutao Zhu},
  \bibinfo{person}{Sirui Wang}, \bibinfo{person}{Fuzheng Zhang},
  \bibinfo{person}{Zhongyuan Wang}, {and} \bibinfo{person}{Ji-Rong Wen}.}
  \bibinfo{year}{2020}\natexlab{}.
\newblock \showarticletitle{S3-rec: Self-supervised learning for sequential
  recommendation with mutual information maximization}. In
  \bibinfo{booktitle}{\emph{Proc. CIKM}}. \bibinfo{pages}{1893--1902}.
\newblock


\bibitem[Zimdars et~al\mbox{.}(2001)]%
        {zimdars2001temporal}
\bibfield{author}{\bibinfo{person}{Andrew Zimdars},
  \bibinfo{person}{David~Maxwell Chickering}, {and}
  \bibinfo{person}{Christopher Meek}.} \bibinfo{year}{2001}\natexlab{}.
\newblock \showarticletitle{Using temporal data for making recommendations}. In
  \bibinfo{booktitle}{\emph{Proc. UAI}}. \bibinfo{pages}{580--588}.
\newblock


\end{thebibliography}

\end{document}